\documentclass[a4paper,twocolumn,english,pra, eqsecnum]{revtex4}
\usepackage[T1]{fontenc}
\usepackage[latin1]{inputenc}
\usepackage{amsmath}

\makeatletter

\date{\today}
\usepackage{ae,aecompl}

\usepackage{bm}

\usepackage{babel}
\makeatother
\begin{document}

\title{Gaussian operator bases for correlated fermions}

\author{J. F. Corney and P. D. Drummond}

\affiliation{ARC Centre of Excellence for Quantum-Atom Optics, University of Queensland,
Brisbane 4072, Queensland, Australia.}

\begin{abstract}
We formulate a general multi-mode Gaussian operator basis for fermions,
to enable a positive phase-space representation of correlated Fermi
states. The Gaussian basis extends existing bosonic phase-space methods
to Fermi systems and thus enables first-principles dynamical or equilibrium
calculations in quantum many-body Fermi systems. We prove the completeness
and positivity of the basis, and derive differential forms for products
with one- and two-body operators. Because the basis satisfies fermionic
superselection rules, the resulting phase space involves only c-numbers,
without requiring anti-commuting Grassmann variables.
\end{abstract}
\maketitle

\section{Introduction}

In this paper we address the issue of how to represent highly correlated
fermionic states, for the purposes of efficient calculations in fermionic
many-body physics. To this end, we introduce a normally ordered Gaussian
operator basis for fermionic density operators. With this basis, earlier
phase-space techniques used to represent atomic transitions\cite{Louisell,SmithGardiner88}
can be extended to general Fermi systems. 

The analogous phase-space methods for bosons were first introduced
over a classical phase space\cite{Wig-Wigner,Hus-Q,Gla-P,CG-Q}, which
was subsequently extended to a non-classical phase space\cite{positiveP1,positiveP2}
and then to a complete Gaussian basis \cite{Gauss:Bosons}. Just as
with bosons, the Gaussian basis enables a representation of arbitrary
fermionic density operators as a positive distribution over a generalised
phase space. 

We investigate both number-conserving basis sets, which generalize
the usual thermal states used to describe Fermi gases, and non-number-conserving
basis sets, which generalize the BCS states found in superconductivity
theory\cite{BCS}. The latter case comprises the most general Gaussian
basis.

We concentrate on the foundational issues of the Gaussian representation
method, proving three central results: 

\begin{itemize}
\item the basis is complete, 
\item the distribution can always be chosen positive, 
\item all two-body operators map to a second-order differential form. 
\end{itemize}
From these results, it follows that positive-definite Fokker-Planck
equations\cite{Gardinermethods} exist for many-body fermionic systems.
Such Fokker-Planck equations enable first-principles stochastic numerical
simulation methods, either in real time or at finite temperature.
They can also be used to obtain novel types of perturbation theory
using stochastic diagram techniques\cite{Stochasticdiagram}. 

The formulation of the resultant phase-space simulation methods, which,
for example, can be applied to the Hubbard model\cite{GaussianQMC},
will be given elsewhere\cite{CorneyDrummond05B}. However, we note
these methods have similarities to the auxiliary field method\cite{Rombouts}
used to simplify fermionic path integrals, except that here the Gaussian
basis is used to expand the fermionic states themselves, rather than
a path integral, which has advantages in terms of giving a greater
physical understanding and fewer restrictions in the resulting applications.

To begin, we establish in Sec.~\ref{sec:Definitions} the definition
of a Gaussian operator in unnormalised form, and follow this in Sec.~\ref{sec:One-and-two}
with elementary examples of one- and two-mode Gaussians in order to
illustrate the basic structure and the physics underlying it. The
two-mode examples include the density operators for thermal and squeezed
fermionic states. In Sec.~\ref{sec:Mathematical-definitions}, we
introduce some notation that is convenient for discussing general
Gaussian operators in a multimode system. 

The main part of the paper (Section \ref{sec:Gaussian-operators})
introduces the most general class of normalized Gaussian operators
and discusses those of its properties that are relevant to its use
as a basis set. This section also gives the proofs of the completeness
and positivity properties, which are established by expansion in number-state
projectors. Other properties are proved by use of Fermi coherent states\cite{CahillGlauber99}
(Appendix \ref{sec:Anticommuting-Algebra}), and these latter proofs
are given in Appendix \ref{sec:Gaussian-fermion-operators}.

\section{Definitions}

\label{sec:Definitions}To define Gaussian operators for a given fermionic
system, we first decompose it into a set of $M$ orthogonal single-particle
modes, or orbitals. With each of these modes, we associate creation
and annihilation operators $\widehat{b}_{j}^{\dagger}$ and $\widehat{b}_{j}$,
with anticommutation relations \begin{eqnarray}
\left[\widehat{b}_{j},\widehat{b}_{k}^{\dagger}\right]_{+} & = & \delta_{jk}\nonumber \\
\left[\widehat{b}_{j},\widehat{b}_{k}\right]_{+} & = & 0\,\,,\label{eq:anticommute}\end{eqnarray}
 where $j,k=1...M$. Thus, $\widehat{\bm b}$ is a column vector of
the $M$ annihilation operators, and $\widehat{\bm b}^{\dagger}$
is a row vector of the corresponding creation operators.

We define a Gaussian operator to be a normally ordered, Gaussian form
of annihilation and creation operators. Like a complex number Gaussian,
the operator Gaussian is an exponential of a quadratic form, with
the exponential defined by its series representation. We follow the
standard convention that fermionic normal ordering includes a sign-change
for every operator exchange, so that $:\widehat{b}_{k}\widehat{b}_{j}^{\dagger}:\,=-\widehat{b}_{j}^{\dagger}\widehat{b}_{k}$.
Normal ordering is utilized here because it allows us to most directly
use Fermi coherent state methods\cite{CahillGlauber99} involving
Grassmann algebra\cite{Berezin}\cite{Balantekin}, which are described
in greater detail in the Appendices. We note that from particle-hole
symmetry, it is also possible to obtain an entirely analogous anti-normally
ordered representation, which simply exchanges the roles of particles
and holes in the formalism.

The most general Gaussian form is a cumbersome object to manipulate,
unless products of odd numbers of operators are excluded. Fortunately,
restricting the set of Gaussians to those containing only even products
can be physically justified on the basis of superselection rules for
fermions. Because it is constructed from pairs of operators, this
type of Gaussian operator contains no Grassmann variables\cite{CahillGlauber99}\cite{Grassmannrep}.

In order to relate these quadratic expressions to the usual Dirac
state-vector notation in the examples that follow, we define number
states as:\begin{eqnarray}
\left|\vec{n}\right\rangle  & = & \left|n_{1}\right\rangle \otimes\left|n_{2}\right\rangle \ldots\otimes\left|n_{M}\right\rangle \nonumber \\
 & = & \left(\widehat{b}_{1}^{\dagger}\right)^{n_{1}}\left(\widehat{b}_{2}^{\dagger}\right)^{n_{2}}\ldots\left(\widehat{b}_{M}^{\dagger}\right)^{n_{M}}\left|0\right\rangle _{M},\label{eq:General-ket-expansion}\end{eqnarray}
where $\vec{n}=\left(n_{1},n_{2},\ldots,n_{M}\right)$ is a vector
of integer occupations and where $\left|0\right\rangle _{M}$ is the
$M$-mode fermion vacuum state. 

We also recall some elementary identities which relate quadratic forms
in fermion operators to state projection operators. For individual
operators, one has the well-known identities:\begin{eqnarray}
\widehat{b}_{j} & = & \left|0\right\rangle _{j}\left\langle 1\right|_{j},\nonumber \\
\widehat{b}_{j}^{\dagger} & = & \left|1\right\rangle _{j}\left\langle 0\right|_{j}.\end{eqnarray}
Hence, the elementary identities for quadratic products are as follows,
for $i<j$ :\begin{eqnarray}
\widehat{b}_{i}^{\dagger}\widehat{b}_{i} & = & \left|1\right\rangle _{i}\left\langle 1\right|_{i}\prod_{k\neq i}\hat{1}_{k}\nonumber \\
\widehat{b}_{i}\widehat{b}_{i}^{\dagger} & = & \left|0\right\rangle _{i}\left\langle 0\right|_{i}\prod_{k\neq i}\hat{1}_{k}\nonumber \\
\widehat{b}_{i}^{\dagger}\widehat{b}_{j} & = & \left|1_{i},0_{j}\right\rangle \left\langle 0_{i},1_{j}\right|\prod_{k\neq i,j}\hat{1}_{k}\nonumber \\
\widehat{b}_{j}\widehat{b}_{i} & = & \left|0_{i},0_{j}\right\rangle \left\langle 1_{i},1_{j}\right|\prod_{k\neq i,j}\hat{1}_{k}\,\,,\label{eq:element-idents}\end{eqnarray}
where $\hat{1}_{k}$ is the identity operator for the $k-$th mode. 

Expanding quantum states in terms of an underlying basis set is a
widespread procedure in quantum mechanics. However, using the general
Gaussian operators in this capacity is nonstandard in several respects.
First, one is expanding a quantum density operator over a basis set
that includes mixed as well as pure quantum states. Second, the Gaussian
operator basis states are not orthogonal. Third, we also allow the
basis set to include operators which are not themselves density operators.
These additional degrees of freedom prove useful in obtaining the
requisite completeness properties. We now give the mathematical form
of these basis operators, in unnormalised form; normalized forms $\widehat{\Lambda}_{M}$,
for which $Tr\left(\widehat{\Lambda}_{M}\right)=1$, will be introduced
in Sec.~\ref{sec:Gaussian-operators}.

\subsection{Number-conserving Gaussians}

We first consider the special case obtained when the quadratic is
of form $\widehat{b}_{i}^{\dagger}\widehat{b}_{j}$, so that it is
number-conserving. The most general, number-conserving Gaussian operator
can be written as:

\begin{eqnarray}
\widehat{\Lambda}_{M}^{(u)}(\mathbf{\bm\mu}) & = & :\exp\left[-\widehat{\bm b}^{\dagger}\bm\mu\widehat{\bm b}\right]:\,\,,\nonumber \\
 & = & :\prod_{i,j=1}^{M}\exp\left[-\widehat{b}_{i}^{\dagger}\mu_{ij}\widehat{b}_{j}\right]:\nonumber \\
 & = & :\prod_{i,j=1}^{M}\left[\hat{1}-\widehat{b}_{i}^{\dagger}\mu_{ij}\widehat{b}_{j}\right]:\label{eq:un-norm}\end{eqnarray}
where $\bm\mu$ is a complex $M\times M$ matrix. The last result
follows because normally ordered products in which the same operator
appears more than once are zero, from the anti-commutation relations
in Eq.~(\ref{eq:anticommute}). For the special case that $\bm\mu$
is a hermitian matrix, the Gaussian operator is just the density operator
of a thermal state. If the eigenvalues of $\bm\mu$ are also either
$0$ or $1$, then the Gaussian corresponds to a Slater determinant
(i.e. a product of single-mode number states).

It should be noted here that we do not restrict the Gaussian operators
of this type to be just thermal states. In general we would wish to
consider density matrices that are linear combinations of Gaussian
operators, and these can exist as hermitian, positive-definite operators
even when composed of Gaussians that have neither property.

\subsection{Non-number-conserving Gaussians}

If anomalous products of form $\widehat{b}_{i}\widehat{b}_{j}$ are
included as well, then the most general non-number-conserving Gaussian
is a type of squeezed state:\begin{eqnarray}
\widehat{\Lambda}_{M}^{(u)}(\mathbf{\bm\mu},\mathbf{\bm\xi},\mathbf{\bm\xi}^{+}) & = & :\exp\left[-\widehat{\bm b}^{\dagger}\bm\mu\widehat{\bm b}-\frac{1}{2}\left(\widehat{\bm b}\bm\xi^{+}\widehat{\bm b}+\widehat{\bm b}^{\dagger}\bm\xi\widehat{\bm b}^{\dagger}\right)\right]:\nonumber \\
 & = & \prod_{i>j}\left[1-\xi_{ij}\hat{b}_{i}^{\dagger}\hat{b}_{j}^{\dagger}\right]:\prod_{ij}\left[1-\mu_{ij}\hat{b}_{i}^{\dagger}\hat{b}_{j}\right]:\times\nonumber \\
 &  & \times\prod_{i>j}\exp\left[1-\xi_{ij}^{+}\hat{b}_{i}\hat{b}_{j}\right]\,\,.\label{eq:sq-un-norm}\end{eqnarray}
Here $\bm\xi$, $\bm\xi^{+}$ are complex antisymmetric $M\times M$
matrices. For the special case that $\bm\mu$ is a hermitian matrix
and $\bm\xi^{\dagger}=\bm\xi^{+}$, then the Gaussian operator is
just the density operator for a squeezed thermal state. Normalized
forms $\widehat{\Lambda}_{M}$, for which $Tr\left(\widehat{\Lambda}_{M}\right)=1$,
will be introduced later.

As before, we do not restrict the Gaussian operators to be just the
squeezed thermal states, even though this represents a large and physically
important class of fermionic density operators. By extending the definition
to include all exponentials of quadratic forms, we can obtain a more
useful set, which is a complete basis with a positive-definite representation
for all density operators, as we show in Sec.~\ref{sec:Gaussian-operators}.

\section{One and two mode Gaussian operators}

\label{sec:One-and-two}In this section we give examples of Gaussian
operators in elementary one and two-mode cases. These are sufficient
to illustrate the basic identities and ideas. More general results
will be given in Sec.~\ref{sec:Gaussian-operators}.

\subsection{Single-mode Gaussian operators}

\label{single}

An unnormalised Gaussian operator for a single mode has only one possible
form:\begin{eqnarray}
\widehat{\Lambda}_{1}^{(u)}(\mu) & = & \,:\exp\left[-\mu\widehat{b}^{\dagger}\widehat{b}\right]:\nonumber \\
 & \equiv & \,:\sum_{k=0}^{\infty}\frac{1}{k!}\left(-\mu\widehat{b}^{\dagger}\widehat{b}\right)^{k}:\,,\label{eq:singlemodegaussian}\end{eqnarray}
where the exponential is defined, as indicated, by its series representation.
Just as in the general case of Eq (\ref{eq:un-norm}), the anticommutivity
of the fermionic operators means that only the zeroth and first-order
terms in the expansion contribute to the single mode Gaussian, giving
the simple result:\begin{eqnarray}
\widehat{\Lambda}_{1}^{(u)}(\mu) & = & 1-\mu\widehat{b}^{\dagger}\widehat{b}.\end{eqnarray}

The normalisation of this Gaussian operator is\begin{eqnarray}
N_{1} & \equiv & \mathrm{Tr}\widehat{\Lambda}_{1}\nonumber \\
 & = & 2-\mu.\end{eqnarray}
 Excluding the point $\mu=2$, we define the new parameter $n=(1-\mu)/(2-\mu)$,
which allows us to write the Gaussian in normalised form as

\begin{eqnarray}
\widehat{\Lambda}_{1}(n) & = & (1-n):\exp\left[-\left(2+1/(n-1)\right)\widehat{b}^{\dagger}\widehat{b}\right]:\nonumber \\
 & = & (1-n)\widehat{b}\widehat{b}^{\dagger}+n\widehat{b}^{\dagger}\widehat{b}\nonumber \\
 & = & (1-n)\left|0\left\rangle \right\langle 0\right|+n\left|1\right\rangle \left\langle 1\right|.\label{eq:onemodegauss}\end{eqnarray}
If $n$ is real, then Eq(\ref{eq:onemodegauss}) shows directly that
$\widehat{\Lambda}_{1}(n)$ is the density operator corresponding
to a mixture of number states in the one-mode case. The expectation
value of the fermion occupation $\widehat{n}\equiv\widehat{b}^{\dagger}\widehat{b}$
is just the variable $n$.

\subsubsection{Completeness and positivity}

As two special cases we obtain the number states:\begin{eqnarray}
\left|0\left\rangle \right\langle 0\right| & = & \widehat{\Lambda}_{1}(0)\nonumber \\
\left|1\right\rangle \left\langle 1\right| & = & \widehat{\Lambda}_{1}(1),\label{eq:numberstate}\end{eqnarray}
 which implies that just these two single-mode Gaussians form a complete
basis set for a number-conserving subset of Hilbert space. Because
super-selection rules prohibit superpositions of states differing
by odd numbers of fermions, this is the most general case possible.
Furthermore, we see from Eq (\ref{eq:numberstate}) that the most
general single-mode density matrix can be expanded as a mixture of
Gaussians with \emph{positive} coefficients, since $0\leq n\leq1$:\begin{equation}
\widehat{\rho}=(1-n)\widehat{\Lambda}_{1}(0)+n\widehat{\Lambda}_{1}(1)\,\,.\label{eq:single-mode-mix}\end{equation}
Additionally in the one-mode case, all physical density operators
are also Gaussian operators $\widehat{\rho}=\widehat{\Lambda}_{1}(n)$,
which is another proof of positivity and completeness.

It is clear from these examples that the Gaussian operators are \emph{overcomplete}:
a physical density matrix may be represented by more than one positive
distribution over the Gaussians. The most general single-mode Gaussian
operator, with $n$ complex, provides an even larger overcomplete
basis for physical density matrices, even though such Gaussians are
not always physical density matrices themselves. For example, any
uniform-phase distribution over Gaussian states gives the zero-particle
state:\begin{eqnarray}
\left|0\left\rangle \right\langle 0\right| & = & \int d\phi\widehat{\Lambda}_{1}(re^{i\phi}).\end{eqnarray}
From the usual hole-particle symmetry of fermion states, the single-particle
state is similarly obtained from:\begin{eqnarray}
\left|1\left\rangle \right\langle 1\right| & = & \int d\phi\widehat{\Lambda}_{1}(1-re^{i\phi}).\end{eqnarray}

\subsection{Two-mode number-conserving Gaussian operators}

\label{two}A straightforward extension of the generalized thermal
Gaussian form to two modes gives\begin{eqnarray}
\widehat{\Lambda}_{2}^{(u)}(\mathbf{\bm\mu}) & = & :\exp\left[-\sum_{i,j=1}^{2}\mu_{ij}\widehat{b}_{i}^{\dagger}\widehat{b}_{j}\right]:\nonumber \\
 & = & 1-\sum_{i,j=1}^{2}\mu_{ij}\widehat{b}_{i}^{\dagger}\widehat{b}_{j}+\det\bm\mu\widehat{b}_{1}^{\dagger}\widehat{b}_{2}^{\dagger}\widehat{b}_{2}\widehat{b}_{1},\nonumber \\
\label{eq:twomodegaussian}\end{eqnarray}
where the $\mu_{ij}$ are the elements of the $2\times2$ matrix $\bm\mu$.
The last step follows by explicitly expanding the general result in
Eq (\ref{eq:un-norm}), while taking account of the sign changes during
normal ordering. Again, the series expansion contains all possible
normally ordered nonzero products of the $\widehat{b}_{i}^{\dagger}\widehat{b}_{j}$
pairs. In terms of two-mode number-state projectors, the Gaussian
operator is\begin{eqnarray}
\widehat{\Lambda}_{2}^{(u)}(\mathbf{\bm\mu}) & = & \left|00\right\rangle \left\langle 00\right|+\left(1-\mu_{11}\right)\left|10\right\rangle \left\langle 10\right|\nonumber \\
 &  & +\left(1-\mu_{22}\right)\left|01\right\rangle \left\langle 01\right|\nonumber \\
 &  & +\left(1-\mu_{11}-\mu_{22}+\det\bm\mu\right)\left|11\right\rangle \left\langle 11\right|\nonumber \\
 &  & -\mu_{12}\left|10\right\rangle \left\langle 01\right|-\mu_{21}\left|01\right\rangle \left\langle 10\right|.\label{eq:twomodegaussianprojectors}\end{eqnarray}

\subsubsection{Normalisation and Moments}

Following from Eq.~(\ref{eq:twomodegaussianprojectors}), the normalisation
is\begin{eqnarray}
N_{2} & = & 4-2\mu_{11}-2\mu_{22}+\det\bm\mu.\end{eqnarray}
 Defining the matrix $\bm n=\mathbf{I}-\left(2\mathbf{I}-\bm\mu^{T}\right)^{-1}$,
where $\mathbf{I}$ is the $2\times2$ identity matrix, we can write
the normalised two-mode Gaussian as\begin{eqnarray}
\widehat{\Lambda}_{2}(\mathbf{n}) & = & \det\left[\mathbf{I}-\mathbf{n}\right]:\exp\left[-\widehat{\bm b}^{\dagger}\left(2\mathbf{I}+\left[\mathbf{n}^{T}-\mathbf{I}\right]^{-1}\right)\widehat{\bm b}\right]:\nonumber \\
 & = & \det\left[\mathbf{I}-\mathbf{n}\right]\left|00\right\rangle \left\langle 00\right|\nonumber \\
 &  & +\left(n_{11}(1-n_{22})+n_{12}n_{21}\right)\left|10\right\rangle \left\langle 10\right|\nonumber \\
 &  & +\left(n_{22}(1-n_{11})+n_{12}n_{21}\right)\left|01\right\rangle \left\langle 01\right|\nonumber \\
 &  & +\det\mathbf{n}\left|11\right\rangle \left\langle 11\right|+n_{21}\left|10\right\rangle \left\langle 01\right|+n_{12}\left|01\right\rangle \left\langle 10\right|.\nonumber \\
\label{eq:TwomodeGauss}\end{eqnarray}

If $\mathbf{n}$ is an Hermitian matrix, then the two-mode Gaussian
corresponds to the density matrix of a mixture of states of different
total number, with coherences $n_{12}=n_{21}^{*}$ between states
of the same total number. 

Normally ordered first-order correlations of the Gaussian density
matrices correspond to elements of $\mathbf{n}$: \begin{eqnarray}
\left\langle \widehat{b}_{i}^{\dagger}\widehat{b}_{j}\right\rangle _{\widehat{\Lambda}} & \equiv & \mathrm{Tr}\widehat{b}_{i}^{\dagger}\widehat{b}_{j}\widehat{\Lambda}_{2}=n_{ij},\label{eq:normalcorrelations}\end{eqnarray}
and higher-order correlations reduce to products of first-order averages,
for example, \begin{eqnarray}
\left\langle \widehat{b}_{1}^{\dagger}\widehat{b}_{1}\widehat{b}_{2}^{\dagger}\widehat{b}_{2}\right\rangle _{\widehat{\Lambda}} & = & n_{11}n_{22}-n_{12}n_{21}.\end{eqnarray}
 This kind of factorisation of higher-order correlations could be
taken as the defining characteristic of a Gaussian state, and more
generally, Gaussian operators, rather than the more formal operator
definition given by Eq.~(\ref{eq:twomodegaussian}). In other words,
Gaussian operators have both a Gaussian form and generate Gaussian
statistics. The connection between these two defining features can
be made more explicit by use of a moment-generating function, or characteristic
function, which is considered in Appendix \ref{sec:Gaussian-fermion-operators}.

Because the matrix $\mathbf{n}$ is Hermitian for a Gaussian that
is a density operator, it can be diagonalised, corresponding to a
change of single-particle basis. In this diagonalised basis, since
the coherences are zero, the density operator is a mixture of number
states, totally characterised by average occupation numbers. In other
words, these Gaussian operators correspond to two-mode thermal states.

\subsubsection{Completeness}

We wish to show first that any number-conserving two-mode density
matrix can be expanded in terms of Gaussian operators, and second
that this can be done with positive expansion coefficients. The first
result follows if we can represent all the number-state projectors
between states of the same total number using Gaussian operators.
By inspection of Eq (\ref{eq:TwomodeGauss}) above, we find that:\begin{eqnarray}
\left|00\right\rangle \left\langle 00\right| & = & \widehat{\Lambda}_{2}(\mathbf{0})\nonumber \\
\left|10\right\rangle \left\langle 10\right| & = & \widehat{\Lambda}_{2}(\left(\begin{array}{cc}
1 & 0\\
0 & 0\end{array}\right))\nonumber \\
\left|01\right\rangle \left\langle 01\right| & = & \widehat{\Lambda}_{2}(\left(\begin{array}{cc}
0 & 0\\
0 & 1\end{array}\right))\nonumber \\
\left|11\right\rangle \left\langle 11\right| & = & \widehat{\Lambda}_{2}(\mathbf{I})\nonumber \\
n_{21}\left|10\right\rangle \left\langle 01\right| & = & \widehat{\Lambda}_{2}(\left(\begin{array}{cc}
n_{1} & 0\\
n_{21} & n_{2}\end{array}\right))-\widehat{\Lambda}_{2}(\left(\begin{array}{cc}
n_{1} & 0\\
0 & n_{2}\end{array}\right))\nonumber \\
n_{12}\left|01\right\rangle \left\langle 10\right| & = & \widehat{\Lambda}_{2}(\left(\begin{array}{cc}
n_{1} & n_{12}\\
0 & n_{2}\end{array}\right))-\widehat{\Lambda}_{2}(\left(\begin{array}{cc}
n_{1} & 0\\
0 & n_{2}\end{array}\right))\nonumber \\
 &  & \,\,.\label{eq:numberconservingprojectors}\end{eqnarray}
Thus, the two-mode Gaussians form a complete operator basis for all
number-conserving density matrices.

\subsubsection{Positivity}

Not only is the two-mode Gaussian a complete representation, but it
is also a positive one: any two-mode number-conserving density operator
can be written as a positive distribution over Gaussian operators.
To see this, note that while the expression for the projection operators,
Eq (\ref{eq:numberconservingprojectors}), includes terms with negative
coefficients, the projectors involved are the off-diagonal ones. Since
density matrices are positive-definite, off-diagonal projectors can
only occur in combination with diagonal projectors. We take this into
account in what follows.

Any two-mode density operator can be expanded into number-state projector
operators as follows:\begin{eqnarray}
\widehat{\rho} & = & \sum_{\vec{n}}\sum_{\vec{n}'}\rho_{\vec{n},\vec{n}'}\left|\vec{n}\right\rangle \left\langle \vec{n}'\right|,\end{eqnarray}
where $\vec{n}$ and $\vec{n}'$ are vectors of integer occupation
numbers: $\vec{n}=(n_{1},n_{2})$, $\vec{n}'=(n_{1}',n_{2}')$. Here
$\rho_{\vec{n},\vec{n}'}=0$ if $\sum_{j}n_{j}\neq\sum_{j}n_{j}'$,
because of total-number conservation. Using the relations in Eq.~(\ref{eq:numberconservingprojectors}),
we can write the density operator as\begin{eqnarray}
\widehat{\rho} & = & \sum_{\vec{n}}\frac{1}{2}\rho_{\vec{n},\vec{n}}\left[\widehat{\Lambda}_{2}(\left(\begin{array}{cc}
n_{1} & 2\rho_{(01),(10)}\\
0 & n_{2}\end{array}\right))\right.\nonumber \\
 &  & \left.+\widehat{\Lambda}_{2}(\left(\begin{array}{cc}
n_{1} & 0\\
2\rho_{(10),(01)} & n_{2}\end{array}\right))\right].\label{eq:generic}\end{eqnarray}
Since the diagonal coefficients $\rho_{\vec{n},\vec{n}}$ are positive
and sum to one, the Gaussian operators form the basis of a probabilistic
representation of any two-mode density operator.

While any two-mode number-conserving density matrix can be expanded
in the form in Eq.~(\ref{eq:generic}), there are often more direct
representations. For example, as Eq.~(\ref{eq:superposition}) below
shows, the entangled state $\left|\phi\right\rangle =\alpha\left|10\right\rangle +\beta\left|01\right\rangle $
can be represented by just one term, rather than the four terms that
result from Eq.~(\ref{eq:generic}).

\subsubsection{Correlation and entanglement}

We first note that any uncorrelated product of number state mixtures
can be represented exactly:\begin{eqnarray}
\widehat{\rho}_{n_{1}}\otimes\widehat{\rho}_{n_{2}} & \equiv & \left\{ (1-n_{1})\left|0\left\rangle \right\langle 0\right|+n_{1}\left|1\left\rangle \right\langle 1\right|\right\} \nonumber \\
 &  & \otimes\left\{ (1-n_{2})\left|0\left\rangle \right\langle 0\right|+n_{2}\left|1\left\rangle \right\langle 1\right|\right\} \nonumber \\
 & = & \widehat{\Lambda}_{2}(\left(\begin{array}{cc}
n_{1} & 0\\
0 & n_{2}\end{array}\right)).\label{eq:thermalstates}\end{eqnarray}
As well as these uncorrelated mixtures, the Gaussian basis can also
be used to represent a mixture with correlations between the modes,
this time as a sum (with positive weights) of two terms:\begin{eqnarray}
A\left|00\right\rangle \left\langle 00\right|+B\left|11\right\rangle \left\langle 11\right| & = & A\widehat{\Lambda}_{2}(\mathbf{0})+B\widehat{\Lambda}_{2}(\mathbf{I}).\nonumber \\
\end{eqnarray}

Importantly, superpositions of number states, corresponding to entangled
states, can also be represented, subject to total-number conservation.
For example, the density matrix corresponding to the state $\left|\phi\right\rangle =\alpha\left|10\right\rangle +\beta\left|01\right\rangle $
is\begin{eqnarray}
\left|\phi\right\rangle \left\langle \phi\right| & = & \left|\alpha\right|^{2}\left|10\right\rangle \left\langle 10\right|+\alpha\beta^{*}\left|10\right\rangle \left\langle 01\right|\nonumber \\
 &  & +\alpha^{*}\beta\left|01\right\rangle \left\langle 10\right|+\left|\beta\right|^{2}\left|01\right\rangle \left\langle 01\right|\nonumber \\
 & = & \widehat{\Lambda}_{2}(\left(\begin{array}{cc}
\left|\alpha\right|^{2} & \alpha^{*}\beta\\
\alpha\beta^{*} & \left|\beta\right|^{2}\end{array}\right)).\label{eq:superposition}\end{eqnarray}
Thus the number-conserving Gaussian operators even include entangled
density matrices, making them a powerful tool for representing highly
correlated states.

This property of being able to represent such Bell-like entangled
states with the hermitian subset of Gaussian operators, is different
to the case of Gaussian expansions for bosons. A typical example of
this type of non-classical representation is the positive P-representation\cite{positiveP1,positiveP2},
which must use a non-hermitian basis to obtain a positive distribution
that can represent all two-mode density matrices, such as those that
violate a Bell inequality\cite{Drummond83}.

\subsection{Two-mode squeezed Gaussian operators}

Equation (\ref{eq:twomodegaussian}) does not represent the most general
two-mode Gaussian form, as the quadratic form does not yet include
terms such as $\widehat{b}_{1}\widehat{b}_{2}$. Incorporating such
anomalous products, we can write the most general Gaussian operator
in normalised form as:\begin{eqnarray}
\widehat{\Lambda}_{2}^{(u)}(\mathbf{\bm\mu},\xi,\xi^{+}) & = & :\exp\left[-\sum_{i,j=1}^{2}\mu_{ij}\widehat{b}_{i}^{\dagger}\widehat{b}_{j}-\xi^{+}\widehat{b}_{1}\widehat{b}_{2}-\xi\widehat{b}_{2}^{\dagger}\widehat{b}_{1}^{\dagger}\right]:\nonumber \\
 & = & 1-\sum_{i,j=1}^{2}\mu_{ij}\widehat{b}_{i}^{\dagger}\widehat{b}_{j}-\xi\widehat{b}_{2}^{\dagger}\widehat{b}_{1}^{\dagger}-\xi^{+}\widehat{b}_{1}\widehat{b}_{2}\nonumber \\
 &  & +\left(\det\bm\mu+\xi\xi^{+}\right)\widehat{b}_{1}^{\dagger}\widehat{b}_{2}^{\dagger}\widehat{b}_{2}\widehat{b}_{1},\label{eq:twomodesqueezedgaussian}\end{eqnarray}
where $\xi$ and $\xi^{+}$ are independent complex numbers. The two
additional operator terms in the expansion, $\widehat{b}_{2}^{\dagger}\widehat{b}_{1}^{\dagger}$
and $\widehat{b}_{1}\widehat{b}_{2}$, are projectors between states
of different total number:\begin{eqnarray}
\widehat{b}_{1}\widehat{b}_{2} & = & -\left|00\right\rangle \left\langle 11\right|,\nonumber \\
\widehat{b}_{2}^{\dagger}\widehat{b}_{1}^{\dagger} & = & -\left|11\right\rangle \left\langle 00\right|,\label{eq:Bellprojectors}\end{eqnarray}
which are the kinds of coherences that appear in the density matrices
of two-mode squeezed states.

\subsubsection{Normalisation and Moments}

The normalisation of the squeezed Gaussian is\begin{eqnarray}
N_{2} & = & 4-2\mu_{11}-2\mu_{22}+\det\bm\mu+\xi\xi^{+}.\end{eqnarray}
To incorporate this into a normalised Gaussian, we redefine the $\mathbf{n}$
matrix to be \begin{eqnarray}
\mathbf{n} & = & \mathbf{I}-\frac{N_{2}}{N_{2}-\xi\xi^{+}}\left(2\mathbf{I}-\bm\mu^{T}\right)^{-1},\end{eqnarray}
and introduce rescaled squeezing parameters $m=-\xi/N_{2}$, $m^{+}=-\xi^{+}/N_{2}$.
The normalised form is then\begin{eqnarray}
\widehat{\Lambda}_{2}(\mathbf{n},m,m^{+}) & = & \left(\det\left[\mathbf{I}-\mathbf{n}\right]+mm^{+}\right)\times\nonumber \\
 &  & :\exp\left[\left(\widehat{\bm b}^{\dagger}\widehat{\bm b}^{T}\right)\left(\underline{\underline{I}}-\underline{\underline{\sigma}}^{-1}/2\right)\left(\begin{array}{c}
\widehat{\bm b}\\
\widehat{\bm b}^{\dagger T}\end{array}\right)\right]:,\nonumber \\
\end{eqnarray}
where the $4\times4$ matrices $\underline{\underline{I}}$ and $\underline{\underline{\sigma}}$
are defined to be\begin{eqnarray}
\underline{\underline{I}} & = & \left[\begin{array}{cccc}
-1 & 0 & 0 & 0\\
0 & -1 & 0 & 0\\
0 & 0 & 1 & 0\\
0 & 0 & 0 & 1\end{array}\right],\nonumber \\
\underline{\underline{\sigma}} & = & \left[\begin{array}{cccc}
n_{11}-1 & n_{21} & 0 & m\\
n_{12} & n_{22}-1 & -m & 0\\
0 & -m^{+} & 1-n_{11} & -n_{12}\\
m^{+} & 0 & -n_{21} & 1-n_{22}\end{array}\right].\end{eqnarray}

In terms of number-state projectors, the normalised Gaussian is a
generalisation of the number-conserving case, but with the additional
non-number-conserving projectors:\begin{eqnarray}
\widehat{\Lambda}_{2}(\underline{\underline{\sigma}}) & = & \left(\det\left[\mathbf{I}-\mathbf{n}\right]+mm^{+}\right)\left|00\right\rangle \left\langle 00\right|\nonumber \\
 &  & +\left(n_{11}(1-n_{22})+n_{12}n_{21}-mm^{+}\right)\left|10\right\rangle \left\langle 10\right|\nonumber \\
 &  & +\left(n_{22}(1-n_{11})+n_{12}n_{21}-mm^{+}\right)\left|01\right\rangle \left\langle 01\right|\nonumber \\
 &  & +\left(\det\mathbf{n}+mm^{+}\right)\left|11\right\rangle \left\langle 11\right|\nonumber \\
 &  & +n_{21}\left|10\right\rangle \left\langle 01\right|+n_{12}\left|01\right\rangle \left\langle 10\right|\nonumber \\
 &  & -m\left|11\right\rangle \left\langle 00\right|-m^{+}\left|00\right\rangle \left\langle 11\right|.\label{eq:sqeezednormalisedgaussian}\end{eqnarray}

In addition to the normal fluctuations of Eq.~(\ref{eq:normalcorrelations}),
the squeezed Gaussians also contain anomalous fluctuations, which
are just equal to the new variables $m$ and $m^{+}$:\begin{eqnarray}
\left\langle \widehat{b}_{1}\widehat{b}_{2}\right\rangle _{\widehat{\Lambda}} & = & m,\nonumber \\
\left\langle \widehat{b}_{2}^{\dagger}\widehat{b}_{1}^{\dagger}\right\rangle _{\widehat{\Lambda}} & = & m^{+},\end{eqnarray}
which implies that for $\widehat{\Lambda}_{2}(\mathbf{n},m,m^{+})$
to be a density matrix, $m$ and $m^{+}$ must be complex-conjugate.
The second-order correlation generalises to\begin{eqnarray}
\left\langle \widehat{b}_{1}^{\dagger}\widehat{b}_{1}\widehat{b}_{2}^{\dagger}\widehat{b}_{2}\right\rangle _{\widehat{\Lambda}} & = & n_{11}n_{22}-n_{12}n_{21}+mm^{+},\end{eqnarray}
which again corresponds to the decorrelation that occurs in a state
with Gaussian statistics.

\subsubsection{Completeness}

From Eq.~(\ref{eq:sqeezednormalisedgaussian}), it follows that the
squeezed Gaussians provide a complete two-mode fermionic basis, not
only for the number conserving subspace, but also for all states containing
superpositions of states whose difference in total number is even.
To see this, note that the projectors between number the $\left|00\right\rangle $
and $\left|11\right\rangle $ number states can be written explicitly
in terms of the Gaussian operators as \begin{eqnarray}
m\left|11\right\rangle \left\langle 00\right| & = & \widehat{\Lambda}_{2}(\mathbf{n},-m,0)-\widehat{\Lambda}_{2}(\mathbf{n},0,0),\nonumber \\
m^{+}\left|00\right\rangle \left\langle 11\right| & = & \widehat{\Lambda}_{2}(\mathbf{n},0,-m^{+})-\widehat{\Lambda}_{2}(\mathbf{n},0,0),\nonumber \\
\end{eqnarray}
for any $\mathbf{n}$. These projectors, together with those of Eq.~(\ref{eq:numberconservingprojectors}),
span the complete Hilbert space of density matrices in question.

\subsubsection{Positivity}

Just as for the number-conserving case, we can write any physical
density operator as a positive distribution over Gaussian operators:\begin{eqnarray}
\widehat{\rho} & = & \sum_{\vec{n}}\sum_{\vec{n}'}\rho_{\vec{n},\vec{n}'}\left|\vec{n}\right\rangle \left\langle \vec{n}'\right|\nonumber \\
 & = & \sum_{\vec{n}}\frac{1}{2}\rho_{\vec{n},\vec{n}}\nonumber \\
 &  & \times\left[\widehat{\Lambda}_{2}(\left(\begin{array}{cc}
n_{1} & 2\rho_{(01),(10)}\\
0 & n_{2}\end{array}\right),-2\rho_{(11,00)},\,0)\right.\nonumber \\
 &  & \left.+\widehat{\Lambda}_{2}(\left(\begin{array}{cc}
n_{1} & 0\\
2\rho_{(10),(01)} & n_{2}\end{array}\right),\,0,-2\rho_{(00,11)})\right].\end{eqnarray}
Thus the Gaussians form the basis of a probabilistic representation
of any physical two-mode density operator. The ability to represent
physical density matrices with the Gaussian basis, either as a single
element or by a positive distribution over basis elements, is important
for calculating dynamical simulations of quantum systems via probabilistic
methods.

\subsubsection{Entanglement}

Again, these types of Gaussians can directly correspond to entangled
density matrices, without having to consider expansions over several
elements. But now superpositions of different total number can be
represented. As an example of these additional kinds of physical states
that the squeezed Gaussians can represent, consider the entangled
non-number-conserving superposition state: $\left|\psi\right\rangle =\left[\left|00\right\rangle +\left|11\right\rangle \right]/\sqrt{2}$.
This can be represented by a single Gaussian:\begin{eqnarray}
\left|\psi\right\rangle \left\langle \psi\right| & = & \frac{1}{2}\left[\left|00\right\rangle \left\langle 00\right|+\left|00\right\rangle \left\langle 11\right|+\left|11\right\rangle \left\langle 00\right|+\left|11\right\rangle \left\langle 11\right|\right]\nonumber \\
 & = & \widehat{\Lambda}_{2}(\frac{1}{2}\mathbf{I},-\frac{1}{2},-\frac{1}{2}).\end{eqnarray}
Hence, the more general type of Gaussian basis element considered
here is considerably more powerful for representing entangled and
correlated states than the number-conserving basis set.

\section{Multimode decomposition of Fermi systems}

\label{sec:Mathematical-definitions}

So far we have considered only one- and two-mode systems, to illustrate
the basic physical properties of Gaussian operators. As we saw, the
Gaussian operators could easily be written in terms of number-state
projectors. The power of the Gaussian operators as a basis in its
own right becomes apparent for multimode systems, for which the number-state
basis becomes unusable. 

Before defining the general Gaussian basis for a system with many
degrees of freedom, we define some mathematical notation and conventions
that will be of subsequent use. As before, we define $\widehat{\bm b}$
as a column vector of the $M$ annihilation operators, and $\widehat{\bm b}^{\dagger}$
as a row vector of the corresponding creation operators. We also introduce
an extended column vector of all $2M$ operators: $\underline{\widehat{b}}=(\widehat{\bm b}^{T},\widehat{\bm b}^{\dagger})^{T}$,
with an adjoint row vector defined as $\underline{\widehat{b}}^{\dagger}=(\widehat{\bm b}^{\dagger},\widehat{\bm b}^{T})$.
Writing these out in full, we get:\begin{equation}
\underline{\widehat{b}}=\left(\begin{array}{c}
\widehat{b}_{1}\\
\vdots\\
\widehat{b}_{M}\\
\widehat{b}_{1}^{\dagger}\\
\vdots\\
\widehat{b}_{M}^{\dagger}\end{array}\right),\,\,\,\underline{\widehat{b}}^{\dagger}=\left(\widehat{b}_{1}^{\dagger},...,\widehat{b}_{M}^{\dagger},\widehat{b}_{1},...,\widehat{b}_{M}\right)\,\,.\label{eq:b-vectors}\end{equation}
 Throughout the paper, we print length-$M$ vectors and $M\times M$
matrices in bold type, and index them where necessary with Latin indices:
$j=1,...,M$. Length-$2M$ vectors we denote with an underline and
$2M\times2M$ matrices we denote with a double underline. These extended
vectors and matrices are indexed where necessary with Greek indices:
$\mu=1,...,2M$. Note that an object such as $\widehat{\underline{b}}\,\widehat{\underline{b}}^{\dagger}$is
a $2M\times2M$ matrix:\begin{eqnarray}
\underline{\widehat{b}}\,\underline{\widehat{b}}^{\dagger} & = & \left[\begin{array}{cc}
\widehat{\mathbf{b}}\widehat{\mathbf{b}}^{\dagger} & \widehat{\mathbf{b}}\widehat{\mathbf{b}}^{T}\\
\widehat{\mathbf{b}}^{\dagger T}\widehat{\mathbf{b}}^{\dagger} & \widehat{\mathbf{b}}^{\dagger T}\widehat{\mathbf{b}}^{T}\end{array}\right]\,\,,\label{eq:bb-matrices}\end{eqnarray}
 whereas $\widehat{\underline{b}}^{\dagger}\,\widehat{\underline{b}}\,=\sum_{k}\widehat{b}_{k}^{\dagger}\,\widehat{b}_{k}+\widehat{b}_{k}\widehat{b}_{k}^{\dagger}(=M)$
is a scalar. More general kinds of vectors are denoted with an arrow
notation: $\overrightarrow{\mu}$.

For products of operators, we make use of normal and antinormal ordering
concepts. Normal ordering, denoted by $:\cdots:$ , is defined as
in the bosonic case, with all annihilation operators to the right
of the creation operators, except that each pairwise reordering involved
induces a sign change, e.~g.~$:\widehat{b}_{i}\widehat{b}_{j}^{\dagger}:\,=-\widehat{b}_{j}^{\dagger}\widehat{b}_{i}\,\,$.
The sign changes are necessary so that the anticommuting natures of
the Fermi operators can be accommodated without ambiguity. We define
\emph{anti}normal ordering similarly, and denote it via curly braces:
$\{\widehat{b}_{j}^{\dagger}\widehat{b}_{i}\}=-\widehat{b}_{i}\widehat{b}_{j}^{\dagger}\,\,$.
More generally, we can define nested orderings, in which the outer
ordering does not reorder the inner one. For example, $\{:\widehat{O}\widehat{b}_{j}^{\dagger}:\widehat{b}_{i}\}=-\widehat{b}_{i}\widehat{b}_{j}^{\dagger}:\widehat{O}:\,\,$,
where $\widehat{O}$ is some operator.

For example, the different orderings of pairs are, in block matrix
form, 

\begin{eqnarray}
:\widehat{b}_{\mu}\,\widehat{b}_{\nu}^{\dagger}:\,=-:\widehat{b}_{\nu}^{\dagger}\,\widehat{b}_{\mu}: & \,= & \left[\begin{array}{cc}
-\widehat{b}_{j}^{\dagger}\widehat{b}_{i} & \widehat{b}_{i}\widehat{b}_{j}\\
\widehat{b}_{i}^{\dagger}\widehat{b}_{j}^{\dagger} & \widehat{b}_{i}^{\dagger}\widehat{b}_{j}\end{array}\right]\,\,,\nonumber \\
\left\{ \widehat{b}_{\mu}\,\widehat{{b}}_{\nu}^{\dagger}\right\} =-\left\{ \widehat{b}_{\nu}^{\dagger}\,\widehat{b}_{\mu}\right\}  & = & \left[\begin{array}{cc}
\widehat{b}_{i}\widehat{b}_{j}^{\dagger} & \widehat{b}_{i}\widehat{b}_{j}\\
\widehat{b}_{i}^{\dagger}\widehat{b}_{j}^{\dagger} & -\widehat{b}_{j}\widehat{b}_{i}^{\dagger}\end{array}\right]\,\,.\end{eqnarray}
Note that this convention means that the relation between the two
orderings is \begin{eqnarray}
:\widehat{\underline{b}}\,\widehat{\underline{b}}^{\dagger}:\, & = & \underline{\underline{I}}+\left\{ \widehat{\underline{b}}\,\widehat{\underline{b}}^{\dagger}\right\} ,\end{eqnarray}
where $\underline{\underline{I}}$ is the constant matrix\begin{eqnarray}
\underline{\underline{I}} & \equiv & \left[\begin{array}{cc}
-\bm I & \bm0\\
\bm0 & \bm I\end{array}\right],\label{eq:constantmatrix}\end{eqnarray}
 in which $\mathbf{0}$ and $\mathbf{I}$ are the $M\times M$ zero
and identity matrices, respectively. 

When ordering products that contain a Gaussian operator $\widehat{\Lambda}$
(and later the density operator), we do not change the ordering of
$\widehat{\Lambda}$ itself; the other operators are merely reordered
around it. Thus $\left\{ :\widehat{\Lambda}\,\widehat{b}_{i}^{\dagger}:\,\widehat{b}_{j}\right\} =-\widehat{b}_{j}\,\widehat{b}_{i}^{\dagger}\widehat{\Lambda}\,$.
The different possible quadratic orderings containing a Gaussian operator
can thus be written in matrix form as

\begin{eqnarray}
:\widehat{\underline{{b}}}\,\widehat{\underline{{b}}}^{\dagger}\widehat{\Lambda}: & = & \left[\begin{array}{cc}
-\left(\widehat{\bm b}^{\dagger T}\widehat{\Lambda}\widehat{\bm b}^{T}\right)^{T} & \widehat{\Lambda}\widehat{\bm b}\widehat{\bm b}^{T}\\
\widehat{\bm b}^{\dagger T}\widehat{\bm b}^{\dagger}\widehat{\Lambda} & \widehat{\bm b}^{\dagger T}\widehat{\Lambda}\widehat{\bm b}^{T}\end{array}\right]\,\,,\nonumber \\
\left\{ \widehat{\underline{{b}}}\,\widehat{\underline{{b}}}^{\dagger}\widehat{\Lambda}\right\}  & = & \left[\begin{array}{cc}
\widehat{\bm b}\widehat{\Lambda}\widehat{\bm b}^{\dagger} & \widehat{\bm b}\widehat{\bm b}^{T}\widehat{\Lambda}\\
\widehat{\Lambda}\widehat{\bm b}^{\dagger T}\widehat{\bm b}^{\dagger} & -\left(\widehat{\bm b}\widehat{\Lambda}\widehat{\bm b}^{\dagger}\right)^{T}\end{array}\right]\,\,,\nonumber \\
\left\{ \widehat{\underline{{b}}}\,:\widehat{\underline{{b}}}^{\dagger}\widehat{\Lambda}:\right\}  & = & \left[\begin{array}{cc}
\widehat{\bm b}\widehat{\bm b}^{\dagger}\widehat{\Lambda} & \widehat{\bm b}\widehat{\Lambda}\widehat{\bm b}^{T}\\
-\left(\widehat{\bm b}^{\dagger T}\widehat{\Lambda}\widehat{\bm b}^{\dagger}\right)^{T} & -\left(\widehat{\Lambda}\widehat{\bm b}\widehat{\bm b}^{\dagger}\right)^{T}\end{array}\right]\,\,,\nonumber \\
\left\{ :\widehat{\Lambda}\widehat{\underline{{b}}}:\,\widehat{\underline{{b}}}^{\dagger}\right\}  & = & \left[\begin{array}{cc}
\widehat{\Lambda}\widehat{\bm b}\widehat{\bm b}^{\dagger} & -\left(\widehat{\bm b}\widehat{\Lambda}\widehat{\bm b}^{T}\right)^{T}\\
\widehat{\bm b}^{\dagger T}\widehat{\Lambda}\widehat{\bm b}^{\dagger} & -\left(\widehat{\bm b}\widehat{\bm b}^{\dagger}\widehat{\Lambda}\right)^{T}\end{array}\right].\label{eq:Matrixform}\end{eqnarray}

\section{Normalized Gaussian operators}

\label{sec:Gaussian-operators}

\subsection{Definition}

We define a normalized Gaussian operator $\widehat{\Lambda}$ to be
the most general Gaussian form of \textit{\emph{fermionic}} annihilation
and creation operators, with zero displacement and unit trace. Using
the extended-vector notation, we first write the normally-ordered
Gaussian in an unnormalized form as

\begin{eqnarray}
\widehat{\Lambda}^{(u)}(\bm\mu,\bm\xi,\bm\xi^{+}) & = & :\exp\left[\underline{\widehat{b}}^{\dagger}\left(\underline{\underline{I}}-\underline{\underline{\sigma}}^{-1}/2\right)\underline{\widehat{b}}\right]:\,,\nonumber \\
\label{eq:Gaussbasis_un}\end{eqnarray}
where we have introduced a new extended $2M\times2M$ covariance matrix
$\underline{\underline{\sigma}}$ defined in terms of $\bm\mu,\bm\xi,\bm\xi^{+}$
so that:\begin{equation}
\underline{\underline{\sigma}}^{-1}=2\underline{\underline{I}}+\left[\begin{array}{cc}
\bm\mu & \bm\xi\\
\bm\xi^{+} & -\bm\mu^{T}\end{array}\right]\,.\label{eq:covariance-mapping}\end{equation}

The introduction of the generalized covariance $\underline{\underline{\sigma}}$
allows the matrix to be written in a normalised form, using the results
of Appendix \ref{sec:Gaussian-fermion-operators}, together with an
explicit complex amplitude factor $\Omega$:

\begin{eqnarray}
\widehat{\Lambda}(\overrightarrow{\lambda}) & = & \Omega\mathrm{Pf}\left[\underline{\underline{\sigma_{A}}}\right]:\exp\left[\underline{\widehat{b}}^{\dagger}\left(\underline{\underline{I}}-\underline{\underline{\sigma}}^{-1}/2\right)\underline{\widehat{b}}\right]:\,.\nonumber \\
\label{eq:Gaussbasis}\end{eqnarray}
 The normalisation is one obvious difference with the conventional
complex-number or bosonic Gaussian forms\cite{Gauss:Bosons}. Chosen
to ensure that ${\rm Tr}\,\widehat{\Lambda}=\Omega\,$, it contains
the Pfaffian of an antisymmetric form $\underline{\underline{\sigma_{A}}}$
of the covariance. The choice of anti-symmetrisation is given in Appendix
\ref{sec:Gaussian-fermion-operators}; other choices will lead, in
general, to additional sign factors. Now the square of the Pfaffian
of an antisymmetric matrix is equal to its determinant, and the determinant
of $\underline{\underline{\sigma_{A}}}$ differs from that of $\underline{\underline{\sigma}}$
by a constant sign (see Appendix \ref{sec:Gaussian-fermion-operators}).
Thus $|\mathrm{Pf}\left[\underline{\underline{\sigma_{A}}}\right]|=|\sqrt{\det\left[\underline{\underline{\sigma}}\right]}|$,
and as we shall see, the relative phase between the two does not appear
in later identities. The additional variable $\Omega$ plays the role
of a weighting factor in the expansion that allows us to represent
unnormalised density operators and to introduce `stochastic gauges'
in the drift\cite{DeuarDrummond01,DeuarDrummond02}.

The covariance has a type of generalized Hermitian antisymmetry, which
can be written as $\underline{\underline{\sigma}}=-\underline{\underline{\sigma}}^{+}$,
with the definition that:

\begin{equation}
\left[\begin{array}{cc}
\mathbf{a} & \mathbf{b}\\
\mathbf{c} & \mathbf{d}\end{array}\right]^{+}\equiv\left[\begin{array}{cc}
\mathbf{d} & \mathbf{c}\\
\mathbf{b} & \mathbf{a}\end{array}\right]^{T}\,\,.\label{eq:sym}\end{equation}
It is this generalised antisymmetry that allows the covariance to
be transformed into an explicitly antisymmetric matrix. The covariance
can also be broken down into the physically significant $M\times M$
submatrices $\mathbf{n}$ , $\mathbf{m}$ and $\mathbf{m}^{+}$:\begin{eqnarray}
\underline{\underline{\sigma}} & = & \left[\begin{array}{cc}
\mathbf{n^{T}}-\mathbf{I} & \mathbf{m}\\
\mathbf{m}^{+} & \mathbf{I}-\mathbf{n}\end{array}\right]\,\,,\label{eq:sigma}\end{eqnarray}
Here $\mathbf{n}$ is a complex matrix, which corresponds to the normal
Green's function in many-body terminology, while $\mathbf{m}$ and
$\mathbf{m}^{+}$ are two independent antisymmetric complex matrices
that correspond to anomalous Green's functions, as we will show in
the next section. 

Thus the total parametrization of a general Gaussian operator is\begin{equation}
\overrightarrow{\lambda}=(\Omega,\mathbf{n},\mathbf{m},\mathbf{m}^{+})\,\,,\label{eq:lambda}\end{equation}
consisting of $1+p=1+M(2M-1)$ parameters in all. However, for many
systems, only a subset of all Gaussian operators is required for a
complete representation of the density operator. One important subset
is the set of generalised thermal states, for which $\mathbf{m}=\mathbf{m}^{+}=\mathbf{0}.$
In this case, from Appendix \ref{sec:Gaussian-fermion-operators},
and using the result that $[2\mathbf{I}-\bm\mu]=[\mathbf{I}-\mathbf{n}]^{-1}$,
the normalization factor is $\det\left[\mathbf{I}-\mathbf{n}\right]$.
The normalized thermal Gaussians therefore can be written: 

\begin{eqnarray}
\widehat{\Lambda}(\overrightarrow{\lambda}) & = & \Omega\det\left[\mathbf{I}-\mathbf{n}\right]:\exp\left[-\widehat{\mathbf{b}}^{\dagger}\left(2\mathbf{I}+\left(\mathbf{n}^{T}-\mathbf{I}\right)^{-1}\right)\widehat{\mathbf{b}}\right]:\,\,.\nonumber \\
\label{eq:ThermalGaussian}\end{eqnarray}

In order to use the Gaussian operator basis, we need to make use of
a number of basic results. The proof of many of these can be established,
as we show in Appendix \ref{sec:Gaussian-fermion-operators}, with
Fermi coherent states and Grassmann algebra, the basics of which are
given in Appendix \ref{sec:Anticommuting-Algebra}. However the final
results do not contain any Grassmann variables.

\subsection{Trace Properties}

\label{sub:Traces}

Some basic traces are

\begin{eqnarray}
{\rm Tr}\left[\widehat{\Lambda}\right] & = & \Omega\,\,,\nonumber \\
{\rm Tr}\left[\underline{b}\,\widehat{\Lambda}\right] & = & 0\,\,,\nonumber \\
{\rm Tr}\left[:\underline{\widehat{b}}\,\underline{\widehat{b}}^{\dagger}\,\widehat{\Lambda}:\right] & = & \Omega\underline{\underline{\sigma}}\,\,.\label{eq:Traces1}\end{eqnarray}
 The first of these is the normalisation, proved as theorem 1 in Appendix
\ref{sec:Gaussian-fermion-operators}. That the second is zero follows
from the fact that the Gaussians are constructed from pairs of ladder
operators and thus cannot correspond to a superposition of states
whose total fermion numbers differ by one. The same result holds for
the trace with any odd product. The third trace, proved as theorem
2 in Appendix \ref{sec:Gaussian-fermion-operators}, allows us to
calculate first-order moments. In terms of the $M\times M$ submatrices,
these become:\begin{eqnarray}
{\rm Tr}\left[\widehat{b}_{i}^{\dagger}\widehat{b}_{j}\,\widehat{\Lambda}\right] & = & \Omega n_{ij}\,,\nonumber \\
{\rm Tr}\left[\widehat{b}_{i}\widehat{b}_{j}\,\widehat{\Lambda}\right] & = & \Omega m_{ij}\,,\nonumber \\
{\rm Tr}\left[\widehat{b}_{i}^{\dagger}\widehat{b}_{j}^{\dagger}\,\widehat{\Lambda}\right] & = & \Omega m_{ij}^{+}\,.\label{eq:Traces2}\end{eqnarray}

These results imply that for $\widehat{\Lambda}$ itself to correspond
to a physical density matrix, $\mathbf{n}$ must be a Hermitian matrix,
since $\left\langle \widehat{b}_{i}^{\dagger}\widehat{b}_{j}\right\rangle ^{*}=\left\langle \widehat{b}_{j}^{\dagger}\widehat{b}_{i}\right\rangle $,
and $\mathbf{m}^{+}$ must be the Hermitian conjugate matrix of $\mathbf{m}$,
since $\left\langle \widehat{b}_{i}\widehat{b}_{j}\right\rangle ^{*}=\left\langle \widehat{b}_{j}^{\dagger}\widehat{b}_{i}^{\dagger}\right\rangle $.
Physically, $\mathbf{n}$ gives the number, or normal, correlations,
and $\mathbf{m}$ and $\mathbf{m}^{+}$ give the squeezing, or anomalous,
correlations.

Another restriction on $\widehat{\Lambda}$ being a physical density
matrix that follows from Eq.~(\ref{eq:Traces2}) is that the eigenvalues
of the matrix $\mathbf{n}$ lie in the interval $\left[0,1\right]$,
because of the Pauli exclusion principle for fermions. Furthermore,
the variance in the number correlations is\begin{eqnarray}
\mathrm{var}\left\{ \left\langle \widehat{b}_{i}^{\dagger}\widehat{b}_{j}\right\rangle \right\}  & = & \left\langle \widehat{b}_{i}^{\dagger}\widehat{b}_{j}\right\rangle \left(1-\left\langle \widehat{b}_{i}^{\dagger}\widehat{b}_{j}\right\rangle \right),\label{eq:FermiVariance}\end{eqnarray}
 which implies that if all the eigenvalues of $\mathbf{n}$ are 0
or 1, then $\widehat{\Lambda}$ is a number state, as the variance
in number vanishes. If the eigenvalues are not limited to 0 or 1,
then the $\widehat{\Lambda}$ corresponds to a mixture of number states
in the eigenbasis and is thus a (possibly squeezed) thermal state,
characterised by average occupation numbers $\overline{n}_{j}=\mathrm{eig}_{j}(\mathbf{n})$,
and squeezing matrix $\mathbf{m}$.

A general Gaussian operator will not necessarily fulfill the Hermiticity
condition and thus will not necessarily correspond to a physical state.
However the set of operators that do correspond to physical states
is an important subclass, because the expansion allows these states
to be represented with exact precision. The inclusion of nonphysical
operators in the expansion, on the other hand, makes the Gaussian
basis an (over)complete basis in which to expand a physical density
operator of an arbitrary state (This is proved below for the general
case). The overcompleteness of the Gaussian operators as a basis set
has important implications for representing arbitrary states with
a positive distribution function, a fact that we discuss in detail
elsewhere\cite{CorneyDrummond05B}.

\subsection{Completeness and Positivity}

We next wish to show that the previous results on completeness and
positivity obtained for one and two mode density matrix representations
can be generalized to the multi-mode case. That is, we will prove
that:

\begin{itemize}
\item For any physical density matrix $\hat{\rho}$, a positive set of coefficients
$P_{j}$ exists such that\begin{equation}
\hat{\rho}=\sum_{j}P_{j}\widehat{\Lambda}(\underline{\underline{\sigma}}^{(j)})\,\,.\end{equation}

\end{itemize}
This central result does not rely on utilising the complex amplitudes
$\Omega$, which are part of the most general Gaussian operator. If
these were used, then positivity of the coefficients would be trivial,
since these additional amplitudes could be used to absorb any phase
or sign factors arising in the expansion. Instead, we wish to prove
a much stronger result, that a positive expansion exists without any
additional amplitude factors. This result is analogous to a similar
result known for the positive-P bosonic representation\cite{positiveP1,positiveP2}.

From the number state basis of Eq (\ref{eq:General-ket-expansion}),
the full set of possible fermionic many-body number states is the
set $\left\{ \left|\vec{n}\right\rangle \right\} $ where $\vec{n}$
is varied over all $2^{M}$ possible permutations. This defines a
complete basis of $2^{2M}$number-state projectors for the set of
all fermionic density operators. While not all the number-state projectors
are Hermitian, it is no restriction to use this larger set of operators
as a basis for the density matrices $\hat{\rho}$.

Next, we expand:\begin{eqnarray}
\hat{\rho} & = & \sum_{\vec{n}}\sum_{\vec{m}}\left|\vec{n}\right\rangle \left\langle \vec{n}\right|\hat{\rho}\left|\vec{m}\right\rangle \left\langle \vec{m}\right|\nonumber \\
 & = & \sum_{\vec{n}}\sum_{\vec{m}}\rho_{\vec{n}\vec{m}}\left|\vec{n}\right\rangle \left\langle \vec{m}\right|\nonumber \\
 & = & \sum_{\vec{n}}\sum_{\vec{m}}\hat{\rho}_{\vec{n}\vec{m}}\end{eqnarray}
The positive definiteness of the density operator means in particular
that diagonal density matrix elements are real and positive: $\rho_{\vec{n}\vec{n}}>0$.
It is sufficient for completeness to prove that any elementary fermionic
operator of form $\hat{\rho}_{\vec{n}\vec{m}}=\rho_{\vec{n}\vec{m}}\left|\vec{n}\right\rangle \left\langle \vec{m}\right|$
corresponds to a normalised Gaussian $\widehat{\Lambda}(\overrightarrow{\lambda})$,
apart from a positive scaling factor. The demonstration proceeds by
constructing limiting cases of Gaussians that correspond to each of
the elementary components of the density matrix. As demonstrated in
the one and two mode cases, such expansions are not unique, and generally
one can obtain other more compact representations by combining diagonal
and off-diagonal elements.

To prove the elementary result, we proceed in three steps:

\subsubsection{Diagonal operators}

First, the generic diagonal operator in the number basis is: \begin{eqnarray}
\hat{\rho}_{\vec{n}\vec{n}} & = & \rho_{\vec{n}\vec{n}}\left|\vec{n}\right\rangle \left\langle \vec{n}\right|\nonumber \\
 & = & \rho_{\vec{n}\vec{n}}\prod_{i}\left|n_{i}\right\rangle _{i}\left\langle n_{i}\right|_{i}\,\,,\label{eq:diagonal_numer_expansion}\end{eqnarray}
with a total occupation number \begin{equation}
N=\sum_{i}n_{i}\,\,.\end{equation}
Each diagonal multi-mode projector $\left|\vec{n}\right\rangle \left\langle \vec{n}\right|$
is simply an outer-product of single-mode density matrices. From the
single-mode example in Section \ref{single}, each single-mode density
matrix corresponds to a Gaussian as in Eq(\ref{eq:single-mode-mix}).
Thus, we see that a diagonal projector is exactly equal to a normalized
Gaussian (we suppress the trivial arguments for simplicity):\begin{eqnarray}
\left|\vec{n}\right\rangle \left\langle \vec{n}\right| & =\prod_{i} & \widehat{\Lambda}_{1}\left[n_{i}\right]\nonumber \\
 & = & \widehat{\Lambda}_{M}\left[\mathbf{n}\right]\,,\end{eqnarray}
where the matrix $\mathbf{n}$ is defined as $n_{ij}=n_{i}\delta_{ij}$. 

Because the $\rho_{\vec{n}\vec{n}}$ are real and positive, Eq.~(\ref{eq:diagonal_numer_expansion})
shows that a positive Gaussian expansion exists for all diagonal density
matrices:\begin{equation}
\widehat{\rho}_{\vec{n}\vec{n}}=\rho_{\vec{n}\vec{n}}\widehat{\Lambda}_{M}\left[\mathbf{n}\right]\,\,.\label{eq:diag-multimode}\end{equation}
 In summary, a diagonal multi-mode projector $\left|\vec{n}\right\rangle \left\langle \vec{n}\right|$
is simply an outer-product of single-mode density matrices, and hence
corresponds exactly to a multi-mode normalized Gaussian.

\subsubsection{Generalized thermal operators}

Next, consider off-diagonal projectors in the number-state expansion
that conserve total number, i.e. \begin{equation}
\widehat{\rho}_{\vec{n}\vec{m}}=\rho_{\vec{n}\vec{m}}\left|\vec{n}\right\rangle \left\langle \vec{m}\right|\,,\end{equation}
 for which \begin{equation}
\sum_{i}n_{i}=N=\sum_{i}m_{i}\,\,.\label{eq:number-conservation}\end{equation}
We show that any such component can be written as the limiting form
of a number-conserving Gaussian, up to a positive scaling factor.

Let \begin{equation}
n_{ij}'=\delta_{ij}\min\{ n_{i},m_{i}\},\label{eq:Min-number}\end{equation}
 and define an off-diagonal Gaussian in terms of the diagonal normalized
Gaussian:\begin{equation}
\widehat{\Lambda}^{(o)}(\bm\mu,\mathbf{n}')=\,:\prod_{i\neq j}\left[1-\mu_{ij}\hat{b}_{i}^{\dagger}\hat{b}_{j}\right]\widehat{\Lambda}\left[\mathbf{n}'\right]:\,.\end{equation}

Now for every distinct mode $i$ with $\delta n_{i}=n_{i}-n_{i}'=1$
one can define a corresponding distinct index $j(i)$ with $\delta m_{j}=m_{j}-m_{j}'=1$.
It follows from the minimum condition, Eq(\ref{eq:Min-number}) that
$j(i)\neq i$, and from the number-conservation equation, Eq (\ref{eq:number-conservation}),
we have \begin{equation}
\sum_{i}\delta n_{i}=\sum_{i}\delta m_{i}\,\,.\end{equation}
Similarly, for every distinct pairs of indices $i,i$' with $\delta n_{i}=\delta n_{i'}=1$,
it follows that $j(i')\neq i$, since otherwise $\delta n_{i}=0$.
The mapping therefore generates distinct pairs so that $j(i)\neq j(i')$$\iff i\neq i'$.
Next, we note that this mapping is not a permutation of the set of
modes $i$ with $\delta n_{i}=0$, since if it were the condition
that $j(i')\neq i$ would be violated for some $i'$. Similarly, the
mapping is not a permutation of any subset of these modes. This means
that the only non-vanishing terms in the normalization factor $\det[2\mathbf{I}-\bm\mu]$
are the diagonal terms, which are already normalized.

Proceed by defining the resulting set of $\delta M\leq M$ pairs as
$\sigma=\{ i,j\}$ , and let: \[
\mu_{ij}(\varepsilon,\vec{n})=\sum_{\{ i',j'\}\in\sigma}\frac{\mu\delta_{ii'}\delta_{jj'}}{-\varepsilon}\]
 where $\mu^{\delta M}=\rho_{\vec{n}\vec{m}}$, so that:

\begin{eqnarray}
\widehat{\Lambda}[\bm\mu(\varepsilon),\mathbf{n}'] & = & :\prod_{i\neq j}\left[1-\mu_{ij}\hat{b}_{i}^{\dagger}\hat{b}_{j}\right]\widehat{\Lambda}\left[\mathbf{n}'\right]:\nonumber \\
 & = & :\prod_{\{ i,j\}\in\sigma}\left[1+\frac{\mu\hat{b}_{i}^{\dagger}\hat{b}_{j}}{\varepsilon}\right]\left|\vec{n}'\right\rangle \left\langle \vec{n}'\right|:\nonumber \\
\end{eqnarray}
 Then consider the limit of $\varepsilon\rightarrow0$, so that to
leading order,

\begin{eqnarray}
\widehat{\rho}_{\vec{n}\vec{m}} & = & \mu^{\delta M}:\prod_{\{ i,j\}\in\sigma}\left[\hat{b}_{i}^{\dagger}\hat{b}_{j}\right]\left|\vec{n}'\right\rangle \left\langle \vec{n}'\right|:\,\nonumber \\
 & = & \lim_{\varepsilon\rightarrow0}\varepsilon^{\delta M}\widehat{\Lambda}[\bm\mu(\varepsilon),\mathbf{n}']\label{eq:gen-thermal-expansion}\end{eqnarray}

Again, we see that a positive expansion parameter is obtained.

\subsubsection{Squeezed operators}

Finally, we consider the remaining elements of the density operator
expansion, i.e. those squeezed projectors $\widehat{\rho}_{\vec{n}\vec{m}}$
for which\begin{equation}
N=\sum_{i}n_{i}=\sum_{i}m_{i}+2S\,\,,\label{eq:non-number-conservation}\end{equation}
 where $S$ is an integer denoting the change in the number of fermion
pairs. We suppose that $S>0$, since the case of $S<0$ is obtained
trivially by hermitian conjugation. Let $\widetilde{n}_{i}$ be obtained
from $n_{i}$ by removing $2S$ occupied sites, labeled as successive
pairs $i,j$ belonging to a set $\widetilde{\sigma},$ so that $\widetilde{N}=\sum_{i}\widetilde{n}_{i}=\sum_{i}m_{i}$.
The occupation numbers $\widetilde{n_{i}},m_{i}$ now define a generalised
thermal density matrix component as previously, and hence equate to
a limiting case of a Gaussian operator from Eq (\ref{eq:gen-thermal-expansion})
above. 

Now define a squeezed off-diagonal Gaussian in terms of the thermal
case, which we have already proved has a positive representation:\begin{equation}
\widehat{\Lambda}^{(s)}(\bm\mu(\varepsilon),\bm\xi(\varepsilon),\mathbf{n}')=\prod_{i>j}\left[1-\xi_{ij}\hat{b}_{i}^{\dagger}\hat{b}_{j}^{\dagger}\right]\widehat{\Lambda}\left[\bm\mu(\varepsilon),\mathbf{n}'\right]\,,\end{equation}
where:

\begin{equation}
\xi_{ij}(\varepsilon)=\sum_{\{ i',j'\}\in\widetilde{\sigma}}\frac{\delta_{ii'}\delta_{jj'}}{-\varepsilon}\,.\end{equation}
Then consider the limit of $\varepsilon\rightarrow0$ as before, so
that to leading order,

\begin{eqnarray}
\widehat{\rho}_{\vec{n}\vec{m}} & = & :\prod_{\{ i,j\}\in\tilde{\sigma}}\left[\hat{a}_{i}^{\dagger}\hat{a}_{j}^{\dagger}\right]\left|\vec{n}'\right\rangle \left\langle \vec{m}\right|:\,\nonumber \\
 & = & \lim_{\varepsilon\rightarrow0}\varepsilon^{S+\delta M}\widehat{\Lambda}^{(s)}[\bm\mu(\varepsilon),\bm\xi(\varepsilon),\mathbf{n}']\end{eqnarray}

A positive expansion parameter is obtained here as well, thus completing
the proof.

\subsection{Differential Properties}

In order to use the Gaussian basis in a time-evolution problem, we
need to be able to map the evolution of the density operator onto
an evolution of the expansion coefficients $P_{j}$. To achieve this,
one must be able to write the action of ladder operators on a Gaussian
basis element in differential form.

We can differentiate the Gaussian operators with respect to their
parameters to get\begin{eqnarray}
\frac{d}{d\Omega}\widehat{\Lambda} & = & \frac{1}{\Omega}\widehat{\Lambda}\,,\nonumber \\
\frac{d}{d\underline{\underline{\sigma}}}\widehat{\Lambda} & = & \underline{\underline{\sigma}}^{-1}\widehat{\Lambda}-\underline{\underline{\sigma}}^{-1}\,:\widehat{\underline{b}}\,\widehat{\underline{b}}^{\dagger}\widehat{\Lambda}:\,\underline{\underline{\sigma}}^{-1}\,,\label{eq:GaussianDeriv}\end{eqnarray}
where the matrix derivative is defined as \begin{eqnarray}
\left(\frac{\partial}{\partial\underline{\underline{\sigma}}}\right)_{\mu,\nu} & = & \frac{\partial}{\partial\sigma_{\nu\mu}}\,\,,\label{eq:definederiv}\end{eqnarray}
 i.~e.~involving a transpose. These expressions for the derivative
can be inverted to obtain the important identities:

\begin{eqnarray}
\widehat{\Lambda} & = & \Omega\frac{\partial}{\partial\Omega}\widehat{\Lambda}\,,\nonumber \\
:\widehat{\underline{b}}\,\widehat{\underline{b}}^{\dagger}\widehat{\Lambda}: & = & \underline{\underline{\sigma}}\widehat{\Lambda}-\underline{\underline{\sigma}}\,\frac{\partial\widehat{\Lambda}}{\partial\underline{\underline{\sigma}}}\,\underline{\underline{\sigma}}\,,\label{eq:Matrixidentities}\end{eqnarray}
and thus we can write the normally ordered action of any pair of operators
on a Gaussian as a first-order derivative. As theorems 5 and 6 in
Appendix \ref{sec:Gaussian-fermion-operators} show, there are analogous
identities for antinormally ordered and mixed pairs:

\begin{eqnarray}
\left\{ \widehat{\underline{b}}:\widehat{\underline{b}}^{\dagger}\widehat{\Lambda}:\right\}  & = & -\underline{\underline{\sigma}}\widehat{\Lambda}+\left(\underline{\underline{\sigma}}-\underline{\underline{I}}\right)\frac{\partial\widehat{\Lambda}}{\partial\underline{\underline{\sigma}}}\underline{\underline{\sigma}}\,,\nonumber \\
\left\{ \widehat{\underline{b}}\,\widehat{\underline{b}}^{\dagger}\widehat{\Lambda}\right\}  & = & \left(\underline{\underline{\sigma}}-\underline{\underline{I}}\right)\widehat{\Lambda}-\left(\underline{\underline{\sigma}}-\underline{\underline{I}}\right)\frac{\partial\widehat{\Lambda}}{\partial\underline{\underline{\sigma}}}\,\left(\underline{\underline{\sigma}}-\underline{\underline{I}}\right)\,.\nonumber \\
\label{eq:Matrixidentities2}\end{eqnarray}

For the subset of Gaussian operators that correspond to (generalised)
thermal states, i.~e.~$\mathbf{m}^{+}=\mathbf{m}=\mathbf{0}$, we
obtain a simpler set of differential identities:\begin{eqnarray}
\widehat{\bm b}^{\dagger T}\widehat{\bm b}^{T}\widehat{\Lambda} & = & \bm n\widehat{\Lambda}+\left(\bm I-\bm n\right)\frac{\partial\widehat{\Lambda}}{\partial\bm n}\bm n\,,\nonumber \\
\widehat{\Lambda}\widehat{\bm b}^{\dagger T}\widehat{\bm b}^{T} & = & \bm n\widehat{\Lambda}+\bm n\frac{\partial\widehat{\Lambda}}{\partial\bm n}\left(\bm I-\bm n\right)\,,\nonumber \\
\widehat{\bm b}^{\dagger T}\widehat{\Lambda}\widehat{\bm b}^{T} & = & \left(\bm I-\bm n\right)\widehat{\Lambda}+\left(\bm I-\bm n\right)\frac{\partial\widehat{\Lambda}}{\partial\bm n}\left(\bm I-\bm n\right)\,,\nonumber \\
\left(\widehat{\bm b}\widehat{\Lambda}\widehat{\bm b}^{\dagger}\right)^{T} & = & \bm n\widehat{\Lambda}-\bm n\frac{\partial\widehat{\Lambda}}{\partial\bm n}\bm n\,.\label{eq:thermal_identities}\end{eqnarray}

The action of four ladder operators on a Gaussian operator can be
obtained by applying the previous identities twice. Thus in a Gaussian
expansion, any two-body interaction term can be written as a second-order
differential operator. As we show in detail elsewhere\cite{CorneyDrummond05B},
this central result allows the evolving density operator to be mapped
to a Fokker-Planck equation for the expansion coefficients, thus enabling
a Monte-Carlo sampling of the many-body quantum state.

\section{Conclusion}

In summary, we have introduced here a generalised Gaussian operator
basis, as a means of defining a phase-space representation for correlated
fermionic states. As special cases, the set of Gaussian operators
include the density operators for thermal states and squeezed states,
and thus the physics of the noninteracting Fermi gas and the BCS state
is incorporated into the basis itself. Furthermore, the basis also
includes more general operators, which ensure an overcompleteness
that makes it possible to express any physical density operator as
a probabilistic distribution over the Gaussian operators, without
the need to use Grassmann algebra. 

We have calculated the normalisation and moments, and proved completeness
and positivity for the most general basis, and also for specific subsets,
such as the number-conserving thermal basis. These results mean that
the phase-space representation defined by the Gaussian operators can
be used for first-principles simulations of many-body fermionic systems.
The mapping from the quantum operator to the probabilistic c-number
description is enabled by the Gaussian differential identities that
have been derived here. The application of these identities will be
dealt with elsewhere, when we consider the phase-space representation
in more detail. 

\begin{acknowledgments}
Funding for this research was provided by an Australian Research Council
Centre of Excellence grant.
\end{acknowledgments}
\appendix

\section{Grassmann Algebra}

\label{sec:Anticommuting-Algebra}This appendix introduces the basic
concepts of noncommuting algebra and lists some results pertaining
to Grassmann calculus and Fermi coherent states. These results are
used in Appendix \ref{sec:Gaussian-fermion-operators} to establish
important properties of the Gaussian operators. Proofs and further
discussion of these results can be found in the literature\cite{Balantekin,Berezin,CahillGlauber99}. 

Let $\bm\alpha$ be a vector of $M$ anticommuting (Grassmann) numbers,
i.e. \begin{eqnarray}
\left[\alpha_{i},\alpha_{j}\right]_{+} & = & 0\,\,.\end{eqnarray}
Since $\alpha_{j}^{2}=0$, any function of Grassmann numbers can be
at most linear in any one of its arguments. Thus, for example, the
single mode exponential is\begin{eqnarray}
\exp\left(\alpha_{j}\right) & = & 1+\alpha_{j}\,\,,\end{eqnarray}
and a multimode exponential, e.~g.~$\exp\left(\sum\alpha_{j}\right)$
will be an ordered product of such single-mode exponentials. 

The Grassmann numbers anticommute with all Fermi annihilation and
creation operators, but commute with c-numbers and bosonic operators.

\subsection{Grassmann Calculus}

Differentiation of a single Grassmann variable is defined as\begin{eqnarray}
\frac{\partial}{\partial\alpha_{i}}\alpha_{j} & = & \delta_{ij}\,\,,\end{eqnarray}
with the derivative of products obtained by the Grassmann chain rule:\begin{eqnarray}
\frac{\partial}{\partial\alpha_{j}}\left[f(\bm\alpha)g(\bm\alpha)\right] & = & \left\{ \begin{array}{cc}
\frac{\partial f}{\partial\alpha_{j}}g+f\frac{\partial g}{\partial\alpha_{j}} & ,\,\,\, f\,{\rm even}\\
\frac{\partial f}{\partial\alpha_{j}}g-f\frac{\partial g}{\partial\alpha_{j}} & ,\,\,\, f\,{\rm odd}\,\end{array}\right.\,\,,\end{eqnarray}
i.~e.~the derivative operator also anticommutes.

Grassmann integration is defined to be the same as differentiation,
but written in a different way:\begin{eqnarray}
\int d\alpha_{j} & = & 0\nonumber \\
\int\alpha_{j}\, d\alpha_{j} & = & 1\,\,.\end{eqnarray}
 Multivariate integrals are ordered sequences of single-variable integrations,
which can be written without ambiguity if the integration measure
(as for a derivative) is also taken to be an anticommuting number.
For an integral over all variables in a vector, we define the integration
measure to be ordered in increasing numerical order: $d\bm\alpha\equiv d\alpha_{1}...d\alpha_{N}\,\,$.

Note that integrating a total derivative gives zero:\begin{eqnarray}
\int d\alpha_{j}\frac{\partial}{\partial\alpha_{j}}f & = & 0\,\,.\end{eqnarray}
 This fact, coupled with the product rule, gives a result for partial
integration:\begin{eqnarray}
\int d\alpha_{j}\frac{\partial f}{\partial\alpha_{j}}g & = & \left\{ \begin{array}{cc}
-\int d\alpha_{j}f\frac{\partial g}{\partial\alpha_{j}} & ,\,\,\, f\,{\rm even}\\
\int d\alpha_{j}f\frac{\partial g}{\partial\alpha_{j}} & ,\,\,\, f\,{\rm odd}\,\end{array}\right.\,\,.\label{eq:grassmann_partial_int}\end{eqnarray}

One very useful result is the multivariate Gaussian integral:\begin{eqnarray}
\int\exp\left(-\bm\alpha^{T}\mathbf{A}\bm\alpha/2\right)d\bm\alpha & = & {\rm Pf}\left[\mathbf{A}\right]\label{eq:grassmann_gaussian_int}\end{eqnarray}
for an antisymmetric matrix $\mathbf{A}$ of complex numbers. The
Pfaffian is related to the determinant $\left({\rm Pf}\left[\mathbf{A}\right]\right)^{2}=\det\mathbf{A}\,\,$,
and thus, apart from a sign change, has many of the same properties.
For example, $\mathrm{Pf}\left[\mathbf{A}^{T}\right]=(-1)^{N}\mathrm{Pf}\left[\mathbf{A}\right]$
and $|\mathrm{Pf}\left[\mathbf{A}^{-1}\right]|=|1/\mathrm{Pf}\left[\mathbf{A}\right]|$.
Another useful type of Gaussian integral is:\begin{eqnarray}
\int\exp\left(-\bm\beta^{T}\mathbf{B}\bm\alpha\right)\prod_{j=1}^{M}\left(d\beta_{j}d\alpha_{j}\right) & = & \det\left[\mathbf{B}\right]\,,\label{eq:grassmann_gaussian_int2}\end{eqnarray}
where \textbf{$\mathbf{B}$} is a square matrix of complex numbers
and $\bm\beta$ and $\bm\alpha$ are two independent Grassmann vectors.
This second integral is in fact a special case of the first, except
with $2M$ total Grassmann variables.

\subsection{Grassmann coherent states}

For each Grassmann number $\alpha_{j}$ we can associate another Grassmann
number, denoted $\overline{\alpha}_{j}\,\,$, to play the role of
a complex conjugate. This is formally regarded as an independent variable
for calculus purposes. The conjugates $\overline{\alpha}_{j}\,\,$
anticommute with all other Grassmann variables. By use of such a complex
Grassmann algebra, we can define a fermionic coherent state, which
is formally an eigenstate of the annihilation operator: \begin{eqnarray}
\left|\alpha_{j}\right\rangle  & = & \left(1+\overline{\alpha}_{j}\alpha_{j}\right)^{-\frac{1}{2}}\left(\left|0\right\rangle +\left|1\right\rangle \alpha_{j}\right)\nonumber \\
 & = & \exp\left(\widehat{b}_{j}^{\dagger}\alpha_{j}-\frac{1}{2}\overline{\alpha}_{j}\alpha_{j}\right)\left|0\right\rangle ,\\
\left\langle \alpha_{j}\right| & = & \left(1+\overline{\alpha}_{j}\alpha_{j}\right)^{-\frac{1}{2}}\left(\left\langle 0\right|+\overline{\alpha}_{j}\left\langle 1\right|\right)\nonumber \\
 & = & \exp\left(\overline{\alpha}_{j}\,\widehat{b}_{j}-\frac{1}{2}\overline{\alpha}_{j}\alpha_{j}\right)\left\langle 0\right|.\label{eq:grassmann_coherent_states}\end{eqnarray}
 Like the bosonic coherent state, the fermionic coherent state can
be written as (Grassmann) displacement from the vacuum:\begin{eqnarray}
\left|\alpha_{j}\right\rangle  & = & \exp\left(\widehat{b}_{j}^{\dagger}\alpha_{j}-\overline{\alpha}_{j}\widehat{b}_{j}\right)\left|0\right\rangle \,\,.\end{eqnarray}
Multimode coherent states are products of the single-mode states:\begin{eqnarray*}
\left|\bm\alpha\right\rangle  & = & \prod_{j=1}^{M}\left|\alpha_{j}\right\rangle \\
 & = & \exp\left[\left(\sum_{j=1}^{M}\widehat{b}_{j}^{\dagger}\alpha_{j}-\overline{\alpha}_{j}\widehat{b}_{j}\right)\right]\left|0\right\rangle \end{eqnarray*}
\\

The inner product of two states is:\begin{eqnarray}
\left\langle \alpha_{i}\right|\left.\!\alpha_{j}\right\rangle  & = & \exp\left(\overline{\alpha}_{i}\alpha_{j}-\overline{\alpha}_{i}\alpha_{i}/2-\overline{\alpha}_{j}\alpha_{j}/2\right)\,\,,\label{eq:Grassmann_inner_product}\end{eqnarray}
and thus as two special cases:\begin{eqnarray}
\left\langle \alpha_{i}\right|\left.\!\alpha_{i}\right\rangle  & = & 1\nonumber \\
\left\langle -\alpha_{i}\right|\left.\!\alpha_{i}\right\rangle  & = & \exp\left(-2\overline{\alpha}_{i}\alpha_{i}\right)\,\,.\end{eqnarray}

The usefulness of the coherent states lies in the fact that they form
a complete set:\begin{eqnarray}
\int d\overline{\alpha}_{j}d\alpha_{j}\left|\alpha_{j}\right\rangle \!\left\langle \alpha_{j}\right| & = & \left|0\right\rangle \!\left\langle 0\right|+\left|1\right\rangle _{j}\!\left\langle 1\right|=I_{1}\,\,,\end{eqnarray}
 or, for the multimode case, \begin{eqnarray}
\int d\underline{\alpha}\left|\bm\alpha\right\rangle \!\left\langle \bm\alpha\right| & = & I_{M}\,\,,\label{eq:grassmann_completeness}\end{eqnarray}
where we have defined a $2M$-variate integration measure as\begin{eqnarray}
d\underline{\alpha} & \equiv & \prod_{j=1}^{M}\left(d\overline{\alpha}_{j}d\alpha_{j}\right).\,\,\end{eqnarray}
Finally, we can express the trace of an arbitrary operator $\widehat{O}$
as\begin{eqnarray}
{\rm Tr}\left[\widehat{O}\right] & = & \int d\underline{\alpha}\left\langle -\bm\alpha\right|\widehat{O}\left|\bm\alpha\right\rangle \,\,.\label{eq:grassmann_trace}\end{eqnarray}

\section{Gaussian fermion operators}

\label{sec:Gaussian-fermion-operators}We prove some useful results
concerning the most general multi-mode Gaussian operator constructed
from fermionic ladder operators. The proofs make use of the properties
of Grassmann coherent states and anticommuting algebra, which are
summarized in Appendix \ref{sec:Anticommuting-Algebra}. However the
final results do not contain any Grassmann variables. The results
establish the trace and differential properties of the Gaussian operators
that are discussed in Sec.~\ref{sec:Gaussian-operators}.

\subsection{General Gaussian Operator}

In this appendix, we use the vector and ordering notations introduced
in Sec.~\ref{sec:Mathematical-definitions}. We find it convenient
to use an unnormalised Gaussian form of Fermi operators:

\begin{eqnarray}
\widehat{\Lambda}^{(u)}(\underline{\underline{\mu}}) & = & :\exp\left[-\underline{\widehat{b}}^{\dagger}\,\underline{\underline{\mu}}\,\underline{\widehat{b}}/2\right]:\,,\nonumber \\
 & \equiv & \sum_{n=0}^{\infty}\frac{1}{2^{n}n!}:\left(-\underline{\widehat{b}}^{\dagger}\,\underline{\underline{\mu}}\,\underline{\widehat{b}}\right)^{n}:\nonumber \\
 & = & :\prod_{\mu,\nu=1}^{2M}\left(1-\frac{1}{2}\widehat{b}_{\mu}^{\dagger}\,\mu_{\mu\nu}\,\widehat{b}_{\nu}\right):\,,\label{eq:Gauss_op_un1}\end{eqnarray}
 where $\underline{\underline{\mu}}$ is a $2M\times2M$ matrix of
parameters, given by\begin{eqnarray*}
\underline{\underline{\mu}} & = & \left[\begin{array}{cc}
\bm\mu & \bm\xi\\
\bm\xi^{+} & -\bm\mu^{T}\end{array}\right]\,.\end{eqnarray*}
In terms of the covariance matrix $\underline{\underline{\sigma}}$,
the unnormalised Gaussian is:

\begin{eqnarray}
\widehat{\Lambda}^{(u)}(\underline{\underline{\sigma}}) & = & :\exp\left[\underline{\widehat{b}}^{\dagger}\left(\underline{\underline{I}}-\underline{\underline{\sigma}}^{-1}/2\right)\underline{\widehat{b}}\right]:\,\,,\label{eq:Gauss_op_un2}\end{eqnarray}
 where the relation between the two parametrizations is $\underline{\underline{\sigma}}=[\underline{\underline{\mu}}+2\underline{\underline{I}}]^{-1}$
and where the diagonal matrix $\underline{\underline{I}}$ is as defined
in Eq.~(\ref{eq:constantmatrix}). 

Because of the anticommuting property of fermions, both $\underline{\underline{\sigma}}$
and $\underline{\underline{\mu}}\,$ possess a generalized antisymmetry:
$\underline{\underline{\sigma}}=-\underline{\underline{\sigma}}^{+}\,\,,\underline{\underline{\mu}}=-\underline{\underline{\mu}}^{+}$,
or more in block matrix form,

\begin{eqnarray}
\left[\begin{array}{cc}
\mathbf{a} & \mathbf{b}\\
\mathbf{c} & \mathbf{d}\end{array}\right] & \equiv & -\underline{\underline{X}}\left[\begin{array}{cc}
\mathbf{a} & \mathbf{b}\\
\mathbf{c} & \mathbf{d}\end{array}\right]^{T}\underline{\underline{X}}\nonumber \\
 & = & -\left[\begin{array}{cc}
\mathbf{d}^{T} & \mathbf{b}^{T}\\
\mathbf{c}^{T} & \mathbf{a}^{T}\end{array}\right]\,\,,\label{eq:sym2}\end{eqnarray}
where the constant matrix $\underline{\underline{X}}$ is defined
as\begin{eqnarray}
\underline{\underline{X}} & \equiv & \left[\begin{array}{cc}
\mathbf{0} & \mathbf{I}\\
\mathbf{I} & \mathbf{0}\end{array}\right].\end{eqnarray}
 When applied to the left of a matrix, $\underline{\underline{X}}$
swaps the upper and lower halves; when applied to the right, it swaps
the left and right halves. This structure means that the matrices
$\underline{\underline{\sigma}}$ and $\underline{\underline{\mu}}$
can be transformed into explicitly antisymmetric forms by certain
permutations of rows and columns. For example, it follows immediately
from Eq.~(\ref{eq:sym2}) that interchanging left and right halves,
or upper and lower halves, will generate an antisymmetric matrix.
Alternatively, inserting each row in the lower half after the corresponding
row in the upper half, and inserting each column in the right half
before the corresponding row in the left half generates the antisymmetric
form\begin{eqnarray}
\left[\begin{array}{cc}
\mathbf{a} & \mathbf{b}\\
\mathbf{c} & \mathbf{d}\end{array}\right]_{A} & \equiv & \left[\begin{array}{ccccc}
b_{11} & a_{11} & \cdots & b_{1M} & a_{1M}\\
d_{11} & c_{11} & \cdots & d_{1M} & c_{1M}\\
\vdots & \vdots & \ddots & \vdots & \vdots\\
b_{M1} & a_{M1} & \cdots & b_{MM} & a_{MM}\\
d_{M1} & c_{M1} & \cdots & d_{MM} & c_{MM}\end{array}\right]\,.\end{eqnarray}
With the covariance matrix antisymmetrized in this way, the Gaussian
operator becomes

\begin{eqnarray}
\widehat{\Lambda}^{(u)}(\underline{\underline{\sigma}}_{A}) & = & :\exp\left[\underline{\widehat{b}}_{A}^{T}\left(\underline{\underline{I}}_{A}-\underline{\underline{\sigma}}_{A}^{-1}/2\right)\underline{\widehat{b}}_{A}\right]:\,\,,\label{eq:Gauss_op_un3}\end{eqnarray}
 where the vector of operators is now \begin{eqnarray}
\underline{\widehat{b}}_{A} & = & \left(\begin{array}{c}
\widehat{b}_{1}\\
\widehat{b}_{1}^{\dagger}\\
\vdots\\
\widehat{b}_{M}\\
\widehat{b}_{M}^{\dagger}\end{array}\right).\end{eqnarray}

The generalised antisymmetry of $\underline{\underline{\sigma}}$
and $\underline{\underline{\mu}}$ has implications for matrix derivatives.
Because each element of the matrix appears twice, we have $\partial\sigma_{\mu\nu}/\partial\sigma_{\theta\phi}=\delta_{\mu\theta}\delta_{\nu\phi}-X_{\mu\phi}X_{\nu\theta}$,
where $\delta_{\mu\nu}$ is the Kronecker delta function. The extra
terms here appear in the derivative of an inverse:\begin{eqnarray}
\frac{\partial\sigma_{\mu\nu}^{-1}}{\partial\sigma_{\theta\phi}} & = & -\sigma_{\mu\theta}^{-1}\sigma_{\phi\nu}^{-1}+\sigma_{\mu\gamma}^{-1}X_{\gamma\phi}X_{\theta\epsilon}\sigma_{\epsilon\nu}^{-1},\label{eq:inverse_deriv}\end{eqnarray}
 and they also give an additional factor of two in the derivative
of a determinant, thus:\begin{eqnarray}
\frac{d}{d\underline{\underline{\sigma}}}\sqrt{\det\underline{\underline{\sigma}}} & = & \sqrt{\det\underline{\underline{\sigma}}}\,\underline{\underline{\sigma}}^{-1},\label{eq:det-deriv}\end{eqnarray}
where we define the matrix derivative as \begin{eqnarray}
\left(\frac{d}{d\underline{\underline{\sigma}}}\right)_{\mu\nu} & = & \frac{d}{d\sigma_{\nu\mu}}\,,\end{eqnarray}
i.~e.~involving a transpose. The result (\ref{eq:inverse_deriv})
allows us to relate the derivatives with respect to $\underline{\underline{\sigma}}$
and $\underline{\underline{\mu}}$ : \begin{eqnarray}
\frac{df}{d\underline{\underline{\mu}}}=\frac{df}{d\underline{\underline{\sigma}}^{-1}} & = & -\underline{\underline{\sigma}}\,\frac{df}{d\underline{\underline{\sigma}}}\,\underline{\underline{\sigma}}\,.\label{eq:lambda-sigma-deriv}\end{eqnarray}

\subsection{Normalisation}

\hfill{}

\noindent \textbf{\emph{Theorem 1:}} The trace of the unnormalised
Gaussian operator is equal to the Pfaffian of the inverse of the antisymmetrized
covariance, i.~e.~

\begin{equation}
N\equiv{\rm Tr}\left[\widehat{\Lambda}^{(u)}(\underline{\underline{\sigma}})\right]={\rm Pf}\left[\underline{\underline{\sigma}}_{A}^{-1}\right]\,\,,\label{eq:norm}\end{equation}
where Pf stands for Pfaffian.

\hfill{}

\noindent \textbf{\emph{Proof:}} Because the Gaussian is in normally
ordered form, it is straightforward to evaluate the trace with multimode
Grassmann coherent states $\left|\bm\alpha\right\rangle $, using
the Grassmann trace result of Eq.~(\ref{eq:grassmann_trace}):\begin{eqnarray}
{\rm Tr}\left[\widehat{\Lambda}^{(u)}\right] & = & \int\left\langle -\bm\alpha\right|\widehat{\Lambda}^{(u)}\left|\bm\alpha\right\rangle \prod_{j}^{M}\left[d\overline{\alpha}_{j}d\alpha_{j}\right]\nonumber \\
 & = & \int\exp\left[-\underline{\alpha}^{\dagger}\,\underline{\underline{\sigma}}^{-1}\,\underline{\alpha}/2\right]\prod_{j}^{M}\left[-d\overline{\alpha}_{j}d\alpha_{j}\right],\nonumber \\
\label{eq:coherent_grassmann}\end{eqnarray}
where we have made use of the fact that, from the Grassmann inner
product result of Eq.~(\ref{eq:Grassmann_inner_product}), $\left\langle -\bm\alpha\right|\left.\bm\alpha\right\rangle =\exp(-2\overline{\bm\alpha}\bm\alpha)$
. We have also changed variables: $\overline{\alpha}_{j}\rightarrow-\overline{\alpha}_{j}\,\,$,
and introduced $2M$-vectors of Grassmann variables $\underline{\alpha}=(\bm\alpha,(\overline{\bm\alpha})^{T})$
and $\underline{\alpha}^{\dagger}=(\overline{\bm\alpha},\bm\alpha^{T})\,\,$.
Changing to the antisymmetric form of the covariance $\underline{\underline{\sigma}}_{A}$,
and swapping the order of the pairs in the integration measure, we
obtain\begin{eqnarray}
{\rm Tr}\left[\widehat{\Lambda}^{(u)}\right] & = & \int\exp\left[-\underline{\alpha}_{A}^{T}\,\underline{\underline{\sigma}}_{A}^{-1}\,\underline{\alpha}_{A}/2\right]\prod_{j}^{M}\left[d\alpha_{j}d\overline{\alpha}_{j}\right]\,,\nonumber \\
\end{eqnarray}
where the reordered Grassmann vector is $\underline{\alpha}_{A}=(\alpha_{1},\overline{\alpha}_{1},...,\alpha_{M},\overline{\alpha}_{M})^{T}.$
Noting that the arrangement of elements in $\underline{\alpha}_{A}$
matches the order of integration, we can apply the Gaussian integral
result {[}Eq.~\ref{eq:grassmann_gaussian_int}){]}:\begin{eqnarray}
{\rm Tr}\left[\widehat{\Lambda}^{(u)}\right] & = & {\rm Pf}\left[\underline{\underline{\sigma}}_{A}^{-1}\right].\label{eq:grassmann_gaussian}\end{eqnarray}
QED.

\hfill{}

\noindent \textbf{\emph{Corollaries:}}

Now the square of the Pfaffian of a matrix is equal to its determinant.
Thus, to within a $\pm$ sign, the normalisation is determined by
the determinant of the covariance:\begin{eqnarray}
\left({\rm Tr}\left[\widehat{\Lambda}^{(u)}\right]\right)^{2} & = & \frac{1}{\det\left[\underline{\underline{\sigma}}_{A}\right]}=\frac{\left(-1\right)^{M}}{\det\left[\underline{\underline{\sigma}}\right]}=\frac{1}{\det\left[i\underline{\underline{\sigma}}\right]}\,\,,\nonumber \\
\end{eqnarray}
 and thus we may write the Gaussian operator in normalised form as\begin{eqnarray}
\widehat{\Lambda} & = & \pm\sqrt{\det\left[i\underline{\underline{\sigma}}\right]}:\exp\left[\underline{\widehat{b}}^{\dagger}\left(\underline{\underline{I}}-\underline{\underline{\sigma}}^{-1}/2\right)\underline{\widehat{b}}\right]:\,\,,\label{eq:Gauss_op_norm}\end{eqnarray}
where the choice of plus/minus sign is determined by $\sqrt{\det\left[\underline{\underline{\sigma}}_{A}\right]}{\rm Pf}\left[\underline{\underline{\sigma}}_{A}^{-1}\right]$.
This extra sign, or phase, which does not appear in the normalizations
of the familiar complex-number or bosonic Gaussians, fortunately does
not appear in any of the identities needed to make use of the Gaussian
operators as a basis for a phase-space representation. 

A specific case where the determinant appears is for the generalised
thermal Gaussian without squeezing parameters, so that $\mathbf{m}=\mathbf{m}^{+}=0$.
In this case, the normalization follows directly from the second Grassmann
Gaussian integral identity, Eq.~(\ref{eq:grassmann_gaussian_int2}).
Following the same procedure as before, we find that:\begin{eqnarray}
{\rm Tr}\left[\widehat{\Lambda}^{(u)}\right] & = & \det\left[2\bm I-\bm\mu\right]\,\,.\label{eq:gen-therm-norm}\end{eqnarray}

\subsection{First-order moments}

\hfill{}

\noindent \textbf{\emph{Theorem 2:}} The fermionic Gaussian operator
is completely characterised by its first-order moments. In particular,
the covariance matrix corresponds to the first-order moments in normally
ordered trace form, i.~e.~ 

\begin{eqnarray}
{\rm Tr\left[:\underline{\widehat{b}}\,\underline{\widehat{b}}^{\dagger}\widehat{\Lambda}:\right]} & = & \underline{\underline{\sigma}}\,\,,\label{eq:quad_moment}\end{eqnarray}
where $\underline{\widehat{b}}\,\underline{\widehat{b}}^{\dagger}$
is a matrix multiplication of two vectors, resulting in the $2M\times2M$
matrix of Eq.~(\ref{eq:bb-matrices}).

\noindent \textbf{\emph{Proof:}} We proceed as in the proof of Theorem
1, by taking the trace of the unnormalised form using Grassmann coherent
states and then changing variables: $\overline{\alpha}_{j}\rightarrow-\overline{\alpha}_{j}\,\,$:\begin{eqnarray}
{\rm Tr}\left[:\underline{\widehat{b}}\,\underline{\widehat{b}}^{\dagger}\widehat{\Lambda}^{(u)}:\right] & = & \int\left\langle -\bm\alpha\right|:\underline{\widehat{b}}\,\underline{\widehat{b}}^{\dagger}\widehat{\Lambda}^{(u)}:\left|\bm\alpha\right\rangle \prod_{j}^{M}\left[d\overline{\alpha}_{j}d\alpha_{j}\right]\nonumber \\
 & = & \int\underline{\alpha}\,\underline{\alpha}^{\dagger}\exp\left[-\underline{\alpha}^{\dagger}\,\underline{\underline{\sigma}}^{-1}\,\underline{\alpha}/2\right]\prod_{j}^{M}\left[d\alpha_{j}d\overline{\alpha}_{j}\right]\,.\nonumber \\
\end{eqnarray}
Next we put the integral into the form Eq.~(\ref{eq:grassmann_gaussian_int}):
${\rm Tr}\left[:\underline{\widehat{b}}\,\underline{\widehat{b}}^{\dagger}\widehat{\Lambda}^{(u)}:\right]$=\begin{eqnarray}
 &  & \int-\left(\underline{\alpha}^{\dagger T}\,\underline{\alpha}^{T}\right)^{T}\exp\left[-\underline{\alpha}^{\dagger}\,\underline{\underline{\sigma}}^{-1}\,\underline{\alpha}/2\right]\prod_{j}^{M}\left[d\alpha_{j}d\overline{\alpha}_{j}\right]\nonumber \\
 & = & \frac{d}{d\underline{\underline{\sigma}}^{-1}}\int\exp\left[-\underline{\alpha}^{\dagger}\,\underline{\underline{\sigma}}^{-1}\,\underline{\alpha}/2\right]\prod_{j}^{M}\left[d\alpha_{j}d\overline{\alpha}_{j}\right]\,\,,\end{eqnarray}
where in taking the derivative with respect to $\underline{\underline{\sigma}}^{-1}$
we have taken account of the fact that each element of it appears
twice, owing to its generalized antisymmetry. Evaluating the Grassmann
integral, and employing the determinant result , we get \begin{eqnarray}
{\rm Tr}\left[:\underline{\widehat{b}}\,\underline{\widehat{b}}^{\dagger}\widehat{\Lambda}^{(u)}:\right] & = & \frac{d}{d\underline{\underline{\sigma}}^{-1}}(\pm)\sqrt{\det\left[i\underline{\underline{\sigma}}^{-1}\right]}\nonumber \\
 & = & \pm\sqrt{\det\left[i\underline{\underline{\sigma}}^{-1}\right]}\underline{\underline{\sigma}}\,\,.\end{eqnarray}
Finally, dividing through by the normalisation in Eq.~(\ref{eq:norm})
gives the normalised result . QED.

\hfill{}

\noindent \textbf{\emph{Corollaries:}}

We can put Eq.~(\ref{eq:quad_moment}) into a more familiar form
by using the cyclic property of trace:\begin{eqnarray}
{\rm Tr\left[:\underline{\widehat{b}}\,\underline{\widehat{b}}^{\dagger}\widehat{\Lambda}:\right]} & = & {\rm Tr}\left[\begin{array}{cc}
-\left(\widehat{\mathbf{b}}^{\dagger T}\widehat{\Lambda}\widehat{\mathbf{b}}^{T}\right)^{T} & \widehat{\Lambda}\widehat{\mathbf{b}}\widehat{\mathbf{b}}^{T}\\
\widehat{\mathbf{b}}^{\dagger T}\widehat{\mathbf{b}}^{\dagger}\widehat{\Lambda} & \widehat{\mathbf{b}}^{\dagger T}\widehat{\Lambda}\widehat{\mathbf{b}}^{T}\end{array}\right]\nonumber \\
 & = & {\rm Tr}\left[\begin{array}{cc}
-\widehat{\mathbf{b}}\widehat{\mathbf{b}}^{\dagger} & \widehat{\mathbf{b}}\widehat{\mathbf{b}}^{T}\\
\widehat{\mathbf{b}}^{\dagger T}\widehat{\mathbf{b}}^{\dagger} & \left(\widehat{\mathbf{b}}\widehat{\mathbf{b}}^{\dagger}\right)^{T}\end{array}\right]\widehat{\Lambda}\nonumber \\
 & = & \left[\begin{array}{cc}
\mathbf{n}^{T}-\mathbf{I} & \mathbf{m}\\
\mathbf{m}^{+} & \mathbf{I}-\mathbf{n}\end{array}\right]\,\,,\end{eqnarray}
where we have defined the matrix $\mathbf{n}$ for the number, or
normal, moments, and the matrices $\mathbf{m}$ and $\mathbf{m}^{+}$
for the squeezing, or anomalous, moments, as follows:\begin{eqnarray}
\mathbf{n} & = & \left\langle \widehat{\mathbf{b}}^{\dagger T}\widehat{\mathbf{b}}^{T}\right\rangle _{\widehat{\Lambda}}\,\,,\nonumber \\
\mathbf{m} & = & \left\langle \widehat{\mathbf{b}}\widehat{\mathbf{b}}^{T}\right\rangle _{\widehat{\Lambda}}\,\,,\nonumber \\
\mathbf{m}^{+} & = & \left\langle \widehat{\mathbf{b}}^{\dagger T}\widehat{\mathbf{b}}^{\dagger}\right\rangle _{\widehat{\Lambda}}\,\,,\label{eq:correlations}\end{eqnarray}
where $\left\langle \widehat{O}\right\rangle _{\widehat{\Lambda}}\equiv{\rm Tr}\left[\widehat{O}\,\widehat{\Lambda}\right]$.
Thus we can write covariance matrix in terms of the moments as\begin{eqnarray}
\underline{\underline{\sigma}} & = & \left[\begin{array}{cc}
\mathbf{n}^{T}-\mathbf{I} & \mathbf{m}\\
\mathbf{m}^{+} & \mathbf{I}-\mathbf{n}\end{array}\right]\,\,,\label{eq:sigma_moments}\end{eqnarray}
or inverting the relationship,\begin{eqnarray}
\left\langle :\widehat{\underline{b}}\,\widehat{\underline{b}}^{\dagger}:\right\rangle _{\widehat{\Lambda}} & = & \underline{\underline{I}}-\underline{\underline{I}}\,\underline{\underline{\sigma}}\,\underline{\underline{I}}\,.\label{eq:normal_moments}\end{eqnarray}

\subsection{Higher-order moments}

One can calculate higher-order moments along similar lines, i.~e.~by
expanding the trace as a Grassmann integral then converting this into
a higher-order derivative of a determinant. The results of this procedure
in the general case can be written in terms of a moment generating
function.

\hfill{}

\noindent \textbf{\emph{Theorem 3:}} Any even moment of a Gaussian
operator, in normally ordered trace form, can be calculated by means
of the moment generating function\begin{eqnarray}
M(\underline{\underline{\tau}}) & \equiv & \sqrt{\det\left[\underline{\underline{I}}^{2}-\underline{\underline{\sigma}}\,\underline{\underline{\tau}}\right]}\label{eq:MGF}\end{eqnarray}
 as follows \begin{eqnarray}
\mathrm{Tr\left[:\widehat{b}_{\mu_{1}}^{\dagger}\widehat{b}_{\mu_{2}}\cdots\widehat{b}_{\mu_{r-1}}^{\dagger}\widehat{b}_{\mu_{r}}\widehat{\Lambda}:\right]} & = & \frac{\partial}{\partial\tau_{\mu_{1},\mu_{2}}}\cdots\nonumber \\
 &  & \left.\cdots\frac{\partial}{\partial\tau_{\mu_{r-1},\mu_{r}}}M(\underline{\underline{\tau}})\right|_{\underline{\underline{\tau}}=0}.\nonumber \\
\label{eq:even_moments}\end{eqnarray}
\hfill{}

\noindent \textbf{\emph{Proof:}} We start by writing the normally
ordered operator product as a derivative of a normally ordered Gaussian:\begin{eqnarray}
:\widehat{b}_{\mu_{1}}^{\dagger}\widehat{b}_{\mu_{2}}\cdots\widehat{b}_{\mu_{r-1}}^{\dagger}\widehat{b}_{\mu_{r}}: & = & \frac{\partial}{\partial\tau_{\mu_{1},\mu_{2}}}\cdots\frac{\partial}{\partial\tau_{\mu_{r-1},\mu_{r}}}\nonumber \\
 &  & \,\,\,\,\left.:\exp\left[\underline{\widehat{b}}^{+}\,\underline{\underline{\tau}}\,\underline{\widehat{b}}/2\right]:\right|_{\underline{\underline{\tau}}=0},\,\,\,\,\,\,\,\,\end{eqnarray}
where $\underline{\underline{\tau}}$ is a $2M\times2M$ matrix of
complex numbers with the generalised antisymmetry $\underline{\underline{\tau}}=-\underline{\underline{\tau}}^{+}$.
Using this result in Eq.~(\ref{eq:even_moments}) shows that the
moment generating function $M(\underline{\underline{\tau}})$ must
satisfy \begin{eqnarray}
M(\underline{\underline{\tau}}) & = & \mathrm{Tr}\left[:\exp\left(\underline{\widehat{b}}^{+}\,\underline{\underline{\tau}}\,\underline{\widehat{b}}/2\right)\widehat{\Lambda}:\right],\end{eqnarray}
which can be evaluated by writing the trace as a Grassmann integral\begin{eqnarray}
M(\underline{\underline{\tau}}) & =\frac{1}{N} & \int\exp\left[-\underline{\alpha}^{\dagger}\,\left(\underline{\underline{\sigma}}^{-1}-\underline{\underline{\tau}}\right)\,\underline{\alpha}/2\right]\prod_{j}^{M}\left[d\alpha_{j}d\overline{\alpha}_{j}\right]\,.\nonumber \\
\end{eqnarray}
 Using the Gaussian integral result (\ref{eq:grassmann_gaussian_int}),
we get \begin{eqnarray}
M(\underline{\underline{\tau}}) & = & {\rm Pf}\left[\underline{\underline{\sigma}}_{A}^{-1}-\underline{\underline{\tau}}_{A}\right]/{\rm Pf}\left[\underline{\underline{\sigma}}_{A}^{-1}\right].\end{eqnarray}
 Because of the relationship between Pfaffians and determinants, we
can rewrite this as\begin{eqnarray}
M(\underline{\underline{\tau}}) & = & \sqrt{\det\left[\underline{\underline{\sigma}}_{A}^{-1}-\underline{\underline{\tau}}_{A}\right]}/\sqrt{\det\left[\underline{\underline{\sigma}}_{A}^{-1}\right]}\nonumber \\
 & = & \sqrt{\det\left[\underline{\underline{I}}^{2}-\underline{\underline{\sigma}}\,\underline{\underline{\tau}}\right],}\end{eqnarray}
where the sign of the square root is chosen to give a positive result
when $\underline{\underline{\tau}}=0$. QED

\noindent \textbf{\emph{Examples}}:

To calculate the derivatives of the moment generating function, one
makes use of the results for the derivative of a determinant Eq.~(\ref{eq:det-deriv})
and of an inverse Eq.~(\ref{eq:inverse_deriv}). A general second-order
moment is of the form\begin{eqnarray}
\mathrm{Tr}\left[:\widehat{b}_{\mu_{1}}\widehat{b}_{\mu_{2}}^{\dagger}\widehat{b}_{\mu_{3}}\widehat{b}_{\mu_{4}}^{\dagger}\widehat{\Lambda}:\right] & = & \sigma_{\mu_{1}\mu_{2}}\sigma_{\mu_{3}\mu_{4}}-\sigma_{\mu_{1}\mu_{4}}\sigma_{\mu_{3}\mu_{2}}\nonumber \\
 & + & \left(\sigma X\right)_{\mu_{1}\mu_{3}}\left(X\sigma\right)_{\mu_{4}\mu_{2}}.\end{eqnarray}
 Thus the normally ordered number-number correlations are\begin{eqnarray}
\mathrm{Tr}\left[:\widehat{n}_{i}\widehat{n}_{j}:\widehat{\Lambda}\right] & = & n_{ii}n_{jj}-n_{ij}n_{ji}-m_{ij}m_{ij}^{+},\,\,\,\,\end{eqnarray}
 i.~e.~containing the three terms expected for a state with Gaussian
correlations. 

Similarly, a third-order moment is of the form\begin{multline}
\mathrm{Tr}\left[:\widehat{b}_{\mu_{1}}\widehat{b}_{\mu_{2}}^{\dagger}\widehat{b}_{\mu_{3}}\widehat{b}_{\mu_{4}}^{\dagger}\widehat{b}_{\mu_{5}}\widehat{b}_{\mu_{6}}^{\dagger}\widehat{\Lambda}:\right]=\,\,\,\,\,\,\,\,\,\,\,\,\,\,\,\,\,\,\,\,\,\,\,\,\,\,\,\,\,\,\,\,\\
\sigma_{\mu_{1}\mu_{2}}\sigma_{\mu_{3}\mu_{6}}\sigma_{\mu_{5}\mu_{4}}+\sigma_{\mu_{1}\mu_{4}}\sigma_{\mu_{3}\mu_{2}}\sigma_{\mu_{5}\mu_{6}}+\sigma_{\mu_{1}\mu_{6}}\sigma_{\mu_{3}\mu_{4}}\sigma_{\mu_{5}\mu_{2}}\\
-\sigma_{\mu_{1}\mu_{2}}\sigma_{\mu_{3}\mu_{4}}\sigma_{\mu_{5}\mu_{6}}-\sigma_{\mu_{1}\mu_{4}}\sigma_{\mu_{3}\mu_{6}}\sigma_{\mu_{5}\mu_{2}}-\sigma_{\mu_{1}\mu_{6}}\sigma_{\mu_{3}\mu_{2}}\sigma_{\mu_{5}\mu_{4}}\\
-\sigma_{\mu_{1}\mu_{2}}\left(\sigma X\right)_{\mu_{3}\mu_{5}}\left(X\sigma\right)_{\mu_{6}\mu_{4}}-\sigma_{\mu_{3}\mu_{4}}\left(\sigma X\right)_{\mu_{1}\mu_{5}}\left(X\sigma\right)_{\mu_{6}\mu_{2}}\\
-\sigma_{\mu_{5}\mu_{6}}\left(\sigma X\right)_{\mu_{1}\mu_{3}}\left(X\sigma\right)_{\mu_{4}\mu_{2}}+\sigma_{\mu_{1}\mu_{4}}\left(\sigma X\right)_{\mu_{3}\mu_{5}}\left(X\sigma\right)_{\mu_{6}\mu_{2}}\\
-\sigma_{\mu_{1}\mu_{6}}\left(\sigma X\right)_{\mu_{3}\mu_{5}}\left(X\sigma\right)_{\mu_{4}\mu_{2}}+\sigma_{\mu_{3}\mu_{2}}\left(\sigma X\right)_{\mu_{1}\mu_{5}}\left(X\sigma\right)_{\mu_{6}\mu_{4}}\\
+\sigma_{\mu_{3}\mu_{6}}\left(\sigma X\right)_{\mu_{1}\mu_{5}}\left(X\sigma\right)_{\mu_{4}\mu_{2}}-\sigma_{\mu_{5}\mu_{2}}\left(\sigma X\right)_{\mu_{1}\mu_{3}}\left(X\sigma\right)_{\mu_{6}\mu_{4}}\\
+\sigma_{\mu_{5}\mu_{4}}\left(\sigma X\right)_{\mu_{1}\mu_{3}}\left(X\sigma\right)_{\mu_{6}\mu_{2}}\,\,\,.\,\,\,\,\,\,\,\,\,\,\,\,\,\,\,\,\,\,\end{multline}
 Thus the normally ordered triple correlations are:\begin{eqnarray}
\mathrm{Tr}\left[:\widehat{n}_{i}\widehat{n}_{j}\widehat{n}_{k}:\widehat{\Lambda}\right] & = & n_{ii}n_{jj}n_{kk}-n_{ii}\left(n_{jk}n_{kj}+m_{jk}m_{jk}^{+}\right)\nonumber \\
 & + & n_{ij}n_{jk}n_{ki}-n_{jj}\left(n_{ik}n_{ki}+m_{ik}m_{ik}^{+}\right)\nonumber \\
 & + & n_{ji}n_{kj}n_{ik}-n_{kk}\left(n_{ij}n_{ji}+m_{ij}m_{ij}^{+}\right)\nonumber \\
 & + & n_{ij}m_{ik}m_{jk}^{+}+n_{ji}m_{jk}m_{ik}^{+}\nonumber \\
 & + & n_{jk}m_{ji}m_{ki}^{+}+n_{kj}m_{ki}m_{ji}^{+}\nonumber \\
 & + & n_{ki}m_{kj}m_{ij}^{+}+n_{ik}m_{kj}m_{ij}^{+},\end{eqnarray}
again as expected for a state with Gaussian correlations.

\subsection{Normally ordered products}

\hfill{}

\noindent \textbf{\emph{Theorem 4:}} A normally ordered product of
a pair of ladder operators and a Gaussian is equivalent to a first-order
differential operator on the Gaussian, as follows:

\begin{eqnarray}
:\widehat{\underline{b}}\,\widehat{\underline{b}}^{\dagger}\widehat{\Lambda}: & = & \underline{\underline{\sigma}}\widehat{\Lambda}-\underline{\underline{\sigma}}\frac{\partial\widehat{\Lambda}}{\partial\underline{\underline{\sigma}}}\underline{\underline{\sigma}}\,\,.\label{eq:normal_identity}\end{eqnarray}

\hfill{}

\noindent \textbf{\emph{Proof:}} The proof can be established easily
without Grassmann algebra. We first take the derivative of the unnormalised
Gaussian operator:\begin{eqnarray}
\frac{\partial}{\partial\mu_{\mu\nu}}\widehat{\Lambda}^{(u)}(\underline{\underline{\mu}}) & =- & :\widehat{b}_{\mu}^{\dagger}\,\widehat{\Lambda}^{(u)}(\underline{\underline{\mu}})\,\widehat{b}_{\nu}:\,\,.\end{eqnarray}
 We write this as a derivative with respect to the covariance matrix,
using Eq.~(\ref{eq:lambda-sigma-deriv}), and swap the pair of operators,
to give, \begin{eqnarray}
\underline{\underline{\sigma}}\,\frac{d\widehat{\Lambda}^{(u)}(\underline{\underline{\sigma}})}{d\underline{\underline{\sigma}}}\,\underline{\underline{\sigma}} & =- & :\underline{\hat{b}}\,\underline{\hat{b}}^{\dagger}\,\widehat{\Lambda}^{(u)}(\underline{\underline{\sigma}}):\,\,.\end{eqnarray}

Next, we take the derivative of the normalisation, using Eq.~(\ref{eq:det-deriv}):\begin{eqnarray}
\frac{d}{d\underline{\underline{\sigma}}}N & = & \frac{d}{d\underline{\underline{\sigma}}}(\pm)\left(\det\left[i\underline{\underline{\sigma}}\right]\right)^{-\frac{1}{2}}\nonumber \\
 & = & -N\underline{\underline{\sigma}}^{-1}\,.\label{eq:norm_deriv}\end{eqnarray}
 Combining both of these results, we find that the derivative of the
normalised Gaussian is\begin{eqnarray}
\frac{d}{d\underline{\underline{\sigma}}}\widehat{\Lambda} & = & -N^{-2}\widehat{\Lambda}^{(u)}\,\frac{dN}{d\underline{\underline{\sigma}}}+N^{-1}\frac{d}{d\underline{\underline{\sigma}}}\widehat{\Lambda}^{(u)}\nonumber \\
 & = & \underline{\underline{\sigma}}^{-1}\widehat{\Lambda}-\underline{\underline{\sigma}}^{-1}\,:\widehat{\underline{b}}\,\widehat{\underline{b}}^{\dagger}\widehat{\Lambda}:\,\underline{\underline{\sigma}}^{-1},\end{eqnarray}
whose inverse is the required result. QED.

This result can also be proved by use of Grassmann coherent-state
expansions, in similar manner to the proofs below for the products
of different ordering.

\subsection{Mixed products}

\hfill{}

\noindent \textbf{\emph{Theorem 5:}} A product of mixed order of a
pair of ladder operators and a Gaussian is equivalent to a first-order
differential operator on the Gaussian, as follows:

\begin{eqnarray}
\left\{ \widehat{\underline{b}}:\widehat{\underline{b}}^{\dagger}\widehat{\Lambda}:\right\}  & = & -\underline{\underline{\sigma}}\widehat{\Lambda}+\left(\underline{\underline{\sigma}}-\underline{\underline{I}}\right)\frac{\partial\widehat{\Lambda}}{\partial\underline{\underline{\sigma}}}\underline{\underline{\sigma}}\,\,.\label{eq:mixed_identity}\end{eqnarray}

\hfill{}

\noindent \begin{flushleft}\textbf{\emph{Proof:}} We first make use
of the Fermi coherent-state completeness identity  to replace all
ladder operators by Grassmann integrals over coherent projection operators:
$\left\{ \widehat{\underline{b}}:\widehat{\underline{b}}^{\dagger}\widehat{\Lambda}:\right\} =$\begin{align}
 & \int d\underline{\gamma}d\underline{\beta}d\underline{\alpha}d\underline{\epsilon}\left|\bm\gamma\right\rangle \!\left\langle \bm\gamma\right|\left\{ \widehat{\underline{b}}\,:\left|\bm\beta\right\rangle \!\left\langle \bm\beta\right|\widehat{\underline{b}}^{\dagger}\widehat{\Lambda}\left|\bm\alpha\right\rangle \!\left\langle \bm\alpha\right|:\right\} \left|\bm\epsilon\right\rangle \!\left\langle \bm\epsilon\right|\nonumber \\
= & \frac{1}{N}\int d\underline{\gamma}d\underline{\beta}d\underline{\alpha}d\underline{\epsilon}\left|\bm\gamma\right\rangle \!\left\langle \bm\epsilon\right|\exp\left[-\left(\overline{\bm\beta},\bm\alpha\right)\left(\underline{\underline{\mu}}+\underline{\underline{I}}\right)\left(\begin{array}{c}
\bm\alpha\\
\overline{\bm\beta}\end{array}\right)/2\right]\nonumber \\
\times & \left(\begin{array}{c}
\bm\beta\\
\overline{\bm\alpha}\end{array}\right)\left(\overline{\bm\beta},\bm\alpha\right)\exp\left[\overline{\bm\gamma}\bm\beta+\overline{\bm\alpha}\bm\epsilon-\overline{\bm\alpha}\bm\alpha-\overline{\bm\beta}\bm\beta-\frac{1}{2}\overline{\bm\gamma}\bm\gamma-\frac{1}{2}\overline{\bm\epsilon}\bm\epsilon\right]\end{align}
\end{flushleft}

\noindent \begin{flushleft}where we have used the result that the
inner product of two coherent states is, from Eq.~(\ref{eq:Grassmann_inner_product})\begin{eqnarray}
\left\langle \bm\beta\right|\left.\!\bm\alpha\right\rangle  & = & \exp\left[\overline{\bm\beta}\bm\alpha-\overline{\bm\beta}\bm\beta/2-\overline{\bm\alpha}\bm\alpha/2\right]\,\,.\end{eqnarray}
 Next, we employ integration by parts {[}Eq.~(\ref{eq:grassmann_partial_int}){]}
to replace $\left(\begin{array}{c}
\bm\beta\\
\overline{\bm\alpha}\end{array}\right)$ by variables that appear in the Gaussian form:\begin{eqnarray}
 &  & \exp\left[-\left(\overline{\bm\beta},\bm\alpha\right)\left(\underline{\underline{\mu}}+\underline{\underline{I}}\right)\left(\begin{array}{c}
\bm\alpha\\
\overline{\bm\beta}\end{array}\right)/2\right]\nonumber \\
 &  & \times\left[\left(\begin{array}{c}
\bm\beta\\
\overline{\bm\alpha}\end{array}\right)\exp\left(-\overline{\bm\alpha}\bm\alpha-\overline{\bm\beta}\bm\beta\right)\right]\left(\overline{\bm\beta},\bm\alpha\right)\nonumber \\
 & = & \exp\left[-\left(\overline{\bm\beta},\bm\alpha\right)\left(\underline{\underline{\mu}}+\underline{\underline{I}}\right)\left(\begin{array}{c}
\bm\alpha\\
\overline{\bm\beta}\end{array}\right)/2\right]\nonumber \\
 &  & \times\left[\left(\begin{array}{c}
-\partial_{\overline{\bm\beta}}\\
\partial_{\bm\alpha}\end{array}\right)\exp\left(-\overline{\bm\alpha}\bm\alpha-\overline{\bm\beta}\bm\beta\right)\right]\left(\overline{\bm\beta},\bm\alpha\right)\nonumber \\
 & \rightarrow & \exp\left(-\overline{\bm\alpha}\bm\alpha-\overline{\bm\beta}\bm\beta\right)\nonumber \\
 &  & \times\left(\begin{array}{c}
\partial_{\overline{\bm\beta}}\\
-\partial_{\bm\alpha}\end{array}\right)\exp\left[-\left(\overline{\bm\beta},\bm\alpha\right)\left(\underline{\underline{\mu}}+\underline{\underline{I}}\right)\left(\begin{array}{c}
\bm\alpha\\
\overline{\bm\beta}\end{array}\right)/2\right]\left(\overline{\bm\beta},\bm\alpha\right)\nonumber \\
 & = & \exp\left(-\overline{\bm\alpha}\bm\alpha-\overline{\bm\beta}\bm\beta\right)\left[\underline{\underline{I}}\left(\underline{\underline{\mu}}+\underline{\underline{I}}\right)\left(\begin{array}{c}
\bm\alpha\\
\overline{\bm\beta}\end{array}\right)\left(\overline{\bm\beta},\bm\alpha\right)-\underline{\underline{I}}\right]\nonumber \\
 &  & \times\exp\left[-\left(\overline{\bm\beta},\bm\alpha\right)\left(\underline{\underline{\mu}}+\underline{\underline{I}}\right)\left(\begin{array}{c}
\bm\alpha\\
\overline{\bm\beta}\end{array}\right)/2\right]\,\,.\nonumber \\
\end{eqnarray}
We can now express the result as a derivative of the unnormalised
Gaussian operator with respect to $\underline{\underline{\mu}}$~:
\begin{flalign}
 & \left\{ \widehat{\underline{b}}:\widehat{\underline{b}}^{\dagger}\widehat{\Lambda}:\right\} =\frac{1}{N}\int d\underline{\gamma}d\underline{\beta}d\underline{\alpha}d\underline{\epsilon}\left|\bm\gamma\right\rangle \!\left\langle \bm\epsilon\right|\nonumber \\
\times & \exp\left[-\overline{\bm\alpha}\bm\alpha-\overline{\bm\beta}\bm\beta-\frac{1}{2}\overline{\bm\gamma}\bm\gamma-\frac{1}{2}\overline{\bm\epsilon}\bm\epsilon+\overline{\bm\gamma}\bm\beta+\overline{\bm\alpha}\bm\epsilon\right]\nonumber \\
\times & \left[\underline{\underline{I}}\left(\underline{\underline{\mu}}+\underline{\underline{I}}\right)\frac{d}{d\underline{\underline{\mu}}}-\underline{\underline{I}}\right]\exp\left[-\frac{1}{2}\left(\overline{\bm\beta},\bm\alpha\right)\left(\underline{\underline{\mu}}+\underline{\underline{I}}\right)\left(\begin{array}{c}
\bm\alpha\\
\overline{\bm\beta}\end{array}\right)\right]\nonumber \\
= & \frac{1}{N}\left[\underline{\underline{I}}\left(\underline{\underline{\mu}}+\underline{\underline{I}}\right)\frac{d}{d\underline{\underline{\mu}}}-\underline{\underline{I}}\right]\widehat{\Lambda}^{(u)}\,\,.\,\,\,\end{flalign}
\end{flushleft}

\noindent Finally, we can change variables to $\underline{\underline{\sigma}}=[\underline{\underline{\mu}}+2\underline{\underline{I}}]^{-1}$
and use the result  for the derivative of the normalisation:\begin{eqnarray}
\left\{ \widehat{\underline{b}}:\widehat{\underline{b}}^{\dagger}\widehat{\Lambda}:\right\}  & = & \frac{1}{N}\left[-\underline{\underline{I}}\left(\underline{\underline{\sigma}}^{-1}-\underline{\underline{I}}\right)\underline{\underline{\sigma}}\,\frac{d\widehat{\Lambda}^{(u)}}{d\underline{\underline{\sigma}}}\underline{\underline{\sigma}}-\underline{\underline{I}}\widehat{\Lambda}^{(u)}\right]\nonumber \\
 & =- & \underline{\underline{I}}\left(\underline{\underline{\sigma}}^{-1}-\underline{\underline{I}}\right)\underline{\underline{\sigma}}\,\left(\frac{d\widehat{\Lambda}}{d\underline{\underline{\sigma}}}+\frac{\widehat{\Lambda}}{N}\frac{dN}{d\underline{\underline{\sigma}}}\right)\underline{\underline{\sigma}}-\underline{\underline{I}}\widehat{\Lambda}\nonumber \\
 & = & \left(\underline{\underline{\sigma}}-\underline{\underline{I}}\right)\left(\frac{d\widehat{\Lambda}}{d\underline{\underline{\sigma}}}-\widehat{\Lambda}\underline{\underline{\sigma}}^{-1}\right)\underline{\underline{\sigma}}-\underline{\underline{I}}\widehat{\Lambda}\nonumber \\
 & = & -\underline{\underline{\sigma}}\widehat{\Lambda}+\left(\underline{\underline{\sigma}}-\underline{\underline{I}}\right)\frac{d\widehat{\Lambda}}{d\underline{\underline{\sigma}}}\,\underline{\underline{\sigma}}\,\,.\end{eqnarray}
QED.

\subsection{Antinormal products }

\hfill{}

\noindent \textbf{\emph{Theorem 6:}} An antinormally ordered product
of a pair of ladder operators and a Gaussian is equivalent to a first-order
differential operator on the Gaussian, as follows:

\begin{eqnarray}
\left\{ \widehat{\underline{b}}\,\widehat{\underline{b}}^{\dagger}\widehat{\Lambda}\right\}  & = & \left(\underline{\underline{\sigma}}-\underline{\underline{I}}\right)\widehat{\Lambda}-\left(\underline{\underline{\sigma}}-\underline{\underline{I}}\right)\frac{\partial\widehat{\Lambda}}{\partial\underline{\underline{\sigma}}}\left(\underline{\underline{\sigma}}-\underline{\underline{I}}\right)\,\,.\label{eq:antinormal_identity}\end{eqnarray}

\hfill{}

\noindent \textbf{\emph{Proof:}} The proof initially proceeds in the
same manner as for products of mixed ordering. We first insert the
coherent state identity to convert the action of the operators into
integrals over coherent-state projectors: $\left\{ \widehat{\underline{b}}\,\widehat{\underline{b}}^{\dagger}\widehat{\Lambda}\right\} =$\begin{align}
 & \int d\underline{\gamma}d\underline{\beta}d\underline{\alpha}d\underline{\epsilon}\left|\bm\gamma\right\rangle \!\left\langle \bm\gamma\right|\left\{ \widehat{\underline{b}}\,\widehat{\underline{b}}^{\dagger}\left|\bm\beta\right\rangle \!\left\langle \bm\beta\right|\widehat{\Lambda}\left|\bm\alpha\right\rangle \!\left\langle \bm\alpha\right|\right\} \left|\bm\epsilon\right\rangle \!\left\langle \bm\epsilon\right|\nonumber \\
= & \frac{1}{N}\int d\underline{\gamma}d\underline{\beta}d\underline{\alpha}d\underline{\epsilon}\left|\bm\gamma\right\rangle \!\left\langle \bm\epsilon\right|\exp\left[-\left(\overline{\bm\beta},\bm\alpha\right)\left(\underline{\underline{\mu}}+\underline{\underline{I}}\right)\left(\begin{array}{c}
\bm\alpha\\
\overline{\bm\beta}\end{array}\right)/2\right]\nonumber \\
\times & \left(\begin{array}{c}
\bm\beta\\
\overline{\bm\alpha}\end{array}\right)\left(\overline{\bm\alpha},\bm\beta\right)\exp\left[\overline{\bm\gamma}\bm\beta+\overline{\bm\alpha}\bm\epsilon-\overline{\bm\alpha}\bm\alpha-\overline{\bm\beta}\bm\beta-\frac{1}{2}\overline{\bm\gamma}\bm\gamma-\frac{1}{2}\overline{\bm\epsilon}\bm\epsilon\right]\end{align}
 This time, however, we integrate by parts twice:\begin{align}
 & \exp\left[-\left(\overline{\bm\beta},\bm\alpha\right)\left(\underline{\underline{\mu}}+\underline{\underline{I}}\right)\left(\begin{array}{c}
\bm\alpha\\
\overline{\bm\beta}\end{array}\right)/2\right]\nonumber \\
 & \times\left(\begin{array}{c}
\bm\beta\\
\overline{\bm\alpha}\end{array}\right)\left(\overline{\bm\alpha},\bm\beta\right)\exp\left(-\overline{\bm\alpha}\bm\alpha-\overline{\bm\beta}\bm\beta\right)\nonumber \\
= & \exp\left[-\left(\overline{\bm\beta},\bm\alpha\right)\left(\underline{\underline{\mu}}+\underline{\underline{I}}\right)\left(\begin{array}{c}
\bm\alpha\\
\overline{\bm\beta}\end{array}\right)/2\right]\nonumber \\
 & \times\left(\begin{array}{c}
-\partial_{\overline{\bm\beta}}\\
\partial_{\bm\alpha}\end{array}\right)\left(\partial_{\bm\alpha},-\partial_{\overline{\bm\beta}}\right)\exp\left(-\overline{\bm\alpha}\bm\alpha-\overline{\bm\beta}\bm\beta\right)\nonumber \\
\rightarrow & \exp\left(-\overline{\bm\alpha}\bm\alpha-\overline{\bm\beta}\bm\beta\right)\left(\begin{array}{c}
-\partial_{\overline{\bm\beta}}\\
\partial_{\bm\alpha}\end{array}\right)\left(\partial_{\bm\alpha},-\partial_{\overline{\bm\beta}}\right)\nonumber \\
 & \times\exp\left[-\left(\overline{\bm\beta},\bm\alpha\right)\left(\underline{\underline{\mu}}+\underline{\underline{I}}\right)\left(\begin{array}{c}
\bm\alpha\\
\overline{\bm\beta}\end{array}\right)/2\right]\nonumber \\
= & \exp\left(-\overline{\bm\alpha}\bm\alpha-\overline{\bm\beta}\bm\beta\right)\left[-\underline{\underline{I}}+\underline{\underline{I}}\left(\underline{\underline{\mu}}+\underline{\underline{I}}\right)\left(\begin{array}{c}
\bm\alpha\\
\overline{\bm\beta}\end{array}\right)\left(\overline{\bm\beta},\bm\alpha\right)\right]\nonumber \\
 & \times\left(\underline{\underline{\mu}}+\underline{\underline{I}}\right)\underline{\underline{I}}\exp\left[-\left(\overline{\bm\beta},\bm\alpha\right)\left(\underline{\underline{\mu}}+\underline{\underline{I}}\right)\left(\begin{array}{c}
\bm\alpha\\
\overline{\bm\beta}\end{array}\right)/2\right]\end{align}
which is now in a form that we can again express as a derivative of
the unnormalised Gaussian operator with respect to $\underline{\underline{\mu}}$~:
\begin{align}
 & \left\{ \widehat{\underline{b}}\,\widehat{\underline{b}}^{\dagger}\widehat{\Lambda}\right\} =\frac{-1}{N}\int d\underline{\gamma}d\underline{\beta}d\underline{\alpha}d\underline{\epsilon}\left|\bm\gamma\right\rangle \!\left\langle \bm\epsilon\right|\nonumber \\
 & \times\exp\left[-\overline{\bm\alpha}\bm\alpha-\frac{1}{2}\overline{\bm\beta}\bm\beta-\overline{\bm\gamma}\bm\gamma-\frac{1}{2}\overline{\bm\epsilon}\bm\epsilon+\overline{\bm\gamma}\bm\beta+\overline{\bm\alpha}\bm\epsilon\right]\nonumber \\
 & \times\left(\left[\underline{\underline{I}}-\underline{\underline{I}}\left(\underline{\underline{\mu}}+\underline{\underline{I}}\right)\frac{d}{d\underline{\underline{\mu}}}\right]\exp\left[-\frac{1}{2}\left(\overline{\bm\beta},\bm\alpha\right)\left(\underline{\underline{\mu}}+\underline{\underline{I}}\right)\left(\begin{array}{c}
\bm\alpha\\
\overline{\bm\beta}\end{array}\right)\right]\right)\nonumber \\
 & \times\left(\underline{\underline{\mu}}+\underline{\underline{I}}\right)\underline{\underline{I}}\nonumber \\
= & \frac{1}{N}\left[-\underline{\underline{I}}\widehat{\Lambda}^{(u)}+\underline{\underline{I}}\left(\underline{\underline{\mu}}+\underline{\underline{I}}\right)\frac{d\widehat{\Lambda}^{(u)}}{d\underline{\underline{\mu}}}\right]\left(\underline{\underline{\mu}}+\underline{\underline{I}}\right)\underline{\underline{I}}\,\,.\end{align}

\noindent Finally, we change variables to $\underline{\underline{\sigma}}=[\underline{\underline{\mu}}+2\underline{\underline{I}}]^{-1}$and
use the result  for the derivative of the normalisation:\begin{eqnarray}
\left\{ \widehat{\underline{b}}\,\widehat{\underline{b}}^{\dagger}\widehat{\Lambda}:\right\}  & = & \frac{1}{N}\left[\underline{\underline{I}}\widehat{\Lambda}^{(u)}-\underline{\underline{I}}\left(\underline{\underline{I}}-\underline{\underline{\sigma}}^{-1}\right)\underline{\underline{\sigma}}\,\frac{d\widehat{\Lambda}^{(u)}}{d\underline{\underline{\sigma}}}\underline{\underline{\sigma}}\right]\left(\underline{\underline{I}}-\underline{\underline{\sigma}}^{-1}\right)\underline{\underline{I}}\nonumber \\
 & = & \left(\underline{\underline{\sigma}}-\underline{\underline{I}}\right)\widehat{\Lambda}-\left(\underline{\underline{\sigma}}-\underline{\underline{I}}\right)\frac{d\widehat{\Lambda}}{d\underline{\underline{\sigma}}}\left(\underline{\underline{\sigma}}-\underline{\underline{I}}\right)\,\,.\end{eqnarray}
QED.


\begin{thebibliography}{10}
\bibitem{Louisell}W. H. Louisell, \emph{Quantum Statistical Properties of Radiation}
(Wiley, New York, 1973).
\bibitem{SmithGardiner88}A. M. Smith and C. W. Gardiner, Phys. Rev. A \textbf{38}, 4073 (1988). 
\bibitem{Wig-Wigner}E.~P.~Wigner, Phys.~Rev. \textbf{40}, 749 (1932). 
\bibitem{Hus-Q}K.~Husimi, Proc.~Phys.~Math.~Soc.~Jpn. \textbf{22}, 264 (1940). 
\bibitem{Gla-P}R.~J.~Glauber, Phys.~Rev. \textbf{131}, 2766 (1963); E.~C.~G.~Sudarshan,
Phys.~Rev.~Lett. \textbf{10}, 277 (1963). 
\bibitem{CG-Q}K.~E.~Cahill and R.~J.~Glauber, Phys.~Rev. \textbf{177}, 1882
(1969).
\bibitem{positiveP1}S.~Chaturvedi, P.~D.~Drummond, and D.~F.~Walls, J.~Phys.~A
\textbf{10}, L187-192 (1977). 
\bibitem{positiveP2}P.~D.~Drummond and C.~W.~Gardiner, J.~Phys.~A \textbf{13}, 2353
(1980). 
\bibitem{Gauss:Bosons}J. F. Corney and P. D. Drummond, Phys.~Rev.~A \textbf{68}, 063822
(2003).
\bibitem{BCS}J. Bardeen, L. N. Cooper and J. R. Schrieffer, Phys.~Rev.~ \textbf{108},
1175 (1957).
\bibitem{Gardinermethods}C. W. Gardiner, \emph{Handbook of Stochastic Methods} (Springer-Verlag,
Berlin, 1983).
\bibitem{Stochasticdiagram}S. Chaturvedi and P. D. Drummond, Eur. Phys. J. B \textbf{8}, 251
(1999).
\bibitem{Rombouts}S. Rombouts and K. Heyde, Phys. Stat. Sol (B) \textbf{237}, 99 (2003)
\bibitem{GaussianQMC}J. F. Corney and P. D. Drummond, Phys.~Rev.~Lett.~\textbf{93},
260401 (2004).
\bibitem{CorneyDrummond05B}J. F. Corney and P. D. Drummond, cond-mat/0411712.
\bibitem{CahillGlauber99}K. E. Cahill and R. J. Glauber, Phys.~Rev.~A \textbf{59}, 1538 (1999).
\bibitem{Berezin}F. A. Berezin, \emph{The Method of Second Quantization} (Academic
Press, New York and London, 1966).
\bibitem{Balantekin}A. B. Balantekin, `An Introduction to Functional Integral Techniques
in Many-Body Physics', in \emph{Modern Perspectives in Many-Body Physics}
(World Scientific, Singapore, 1994), eds M. P. Das and J. Mahanty,
pp155-169.
\bibitem{Grassmannrep}L. I. Plimak, M. J. Collett and M. K. Olsen, Phys.~Rev.~A \textbf{64},
063409 (2001).
\bibitem{Drummond83}P. D. Drummond, Phys. Rev. Lett. \textbf{50}, 1407 (1983). 
\bibitem{DeuarDrummond02}P. Deuar and P. D. Drummond, Phys.~Rev.~A \textbf{66}, 033812 (2002).
\bibitem{DeuarDrummond01}P. Deuar and P. D. Drummond, Comp.~Phys.~Commun.~\textbf{142},
442 (2001).
\end{thebibliography}
\end{document}